\documentclass[twocolumn,superscriptaddress]{revtex4}

\usepackage{graphicx}                   

\usepackage{threeparttable} 
\usepackage{array}

\usepackage{color}                       
\usepackage{float}                       
\usepackage[hidelinks]{hyperref}         
\usepackage{wrapfig}                     

\usepackage[parfill]{parskip}            

\usepackage{amstext,amsmath,amssymb,amsfonts,braket}                 
\usepackage{siunitx}

\usepackage{xspace}

\usepackage[normalem]{ulem}
\usepackage{soul}

\usepackage[defaultcolor=blue]{changes}

\newcommand{\beq}{\begin{equation}}
\newcommand{\eeq}{\end{equation}}

\newcommand{\beqq}{\begin{equation*}}
\newcommand{\eeqq}{\end{equation*}}

\newcommand{\bcen}{\begin{center}}
\newcommand{\ecen}{\end{center}}

\newcommand{\tsc}{\textsc}

\newcommand{\Chi}{\mathcal{X}\xspace}

\renewcommand\hl[1]{%
	\bgroup
	\hskip0pt\color{red}%
	#1%
	\egroup
}

\begin{document}

\title{Signature of attochemical quantum interference upon ionization and excitation of an electronic wavepacket in fluoro-benzene}

\author{Anthony Fert\'e}
\affiliation{Nantes Universit\'e, CNRS, CEISAM, UMR 6230, F-44000 Nantes, France}
\author{Dane Austin}
\author{Allan S. Johnson}
\author{Felicity McGrath}
\affiliation{Quantum Optics and Laser Science Group, Blackett Laboratory, Imperial College London, London, UK}
\author{Jo\~ao Pedro Malhado}
\affiliation{Chemistry Department, Imperial College London, Prince Consort Road, London, SW7 2AZ, UK}
\author{Jon P. Marangos}
\affiliation{Quantum Optics and Laser Science Group, Blackett Laboratory, Imperial College London, London, UK}
\author{Morgane Vacher}\email{morgane.vacher@univ-nantes.fr}
\affiliation{Nantes Universit\'e, CNRS, CEISAM, UMR 6230, F-44000 Nantes, France}

\date{\today}

\begin{abstract}
Ultrashort pulses can excite or ionize molecules and populate coherent electronic wavepackets, inducing complex dynamics. In this work, we simulate the coupled electron-nuclear dynamics upon ionization to different electronic wavepackets of (deuterated) benzene and fluoro-benzene molecules, quantum mechanically and in full dimensionality. In fluoro-benzene, the calculations unravel both inter-state and intra-state quantum interferences that leave clear signatures of attochemistry and charge-directed dynamics in the shape of the autocorrelation function. The latter are in agreement with experimental high harmonic spectroscopy measurements of benzenes and fluoro-benzene.
\end{abstract}
\maketitle

With ultrashort light pulses emerging~\cite{Hentschel-2001,Paul-2001}, the time resolution of measurements is pushed to the intrinsic timescale of electron dynamics in molecules \textit{i.e.}~the attosecond~\cite{Calegari-2014,Kraus-2015,Barillot-2021,Mansson-2021}. Exploiting the large pulse bandwidth to coherently excite several electronic states, and forming a so-called electronic wavepacket~\cite{Cederbaum-1999,Remacle-2006,Nisoli-2017}, unveiled the prospect to go beyond the possibilities of standard photochemistry. 
Coherent superpositions of electronic states can also be produced via strong field ionization~\cite{Smirnova-2009,Goulielmakis-2010}. In both cases, the interference between the components of an electronic wavepacket
 can in principle be used to steer chemical reactivity. Such attochemical control, also called charge-directed reactivity, was demonstrated theoretically and experimentally in diatomics~\cite{Roudnev-2004,Kling-2006}. Extending it to polyatomic molecules is one of the current main prospects of attochemistry~\cite{Merritt-2021,FerVac-INC-22}.

Experimentally, generating and observing such dynamics is challenging. Attosecond pump-probe schemes, although becoming possible~\cite{Travers-2019,Duris-2020,ONeal-2020,Kretschmar-2022}, are arduous. 
As an alternative, high-harmonic generation spectroscopy (HHS)~\cite{Marangos-2016} has been successfully applied to probe the sub-femtosecond coupled electron-nuclear dynamics on simple systems, retrieving autocorrelation functions~\cite{Mairesse-2003,Lein-2005,Baker-2006,Farrell-2011,Diveki-2012,Kraus-2013,Lan-2017,Goncalves-2021}. Its main disadvantage is the presence of a strong field that may perturb the dynamics of interest, although field-free simulations succeeded in reproducing experimental observables~\cite{Lein-2005,Baker-2006}. Recent developments of HHS allowed to measure the dynamics in polyatomic molecules~\cite{Austin-2021}.

Simulating a chemical reaction induced by an electronic wavepacket is also challenging~\cite{Merritt-2021,Tran-2024}. Such works were performed with full quantum dynamics methods and focused on small molecules~\cite{Nikodem-2017,Schnappinger2021} or medium-size molecules in reduced-dimensionality~\cite{Schuppel2020,Valentini2020}. The models used are mostly suited to describe rigid motions and are thus difficult to apply to reactions such as dissociations or \textit{cis}-\textit{trans} isomerisations.  Also, reduced dimensionality may not capture all relevant features of the dynamics.
Other simulations employed mixed quantum-classical dynamics methods~\cite{Meisner-2015,Vacher-2016-FD,Tran-2021,Danilov-2022}, 
not treating electronic coherence --  a key property for attochemistry -- accurately~\cite{Tran-2024}.

\begin{figure}[b]

\includegraphics[scale=0.28]{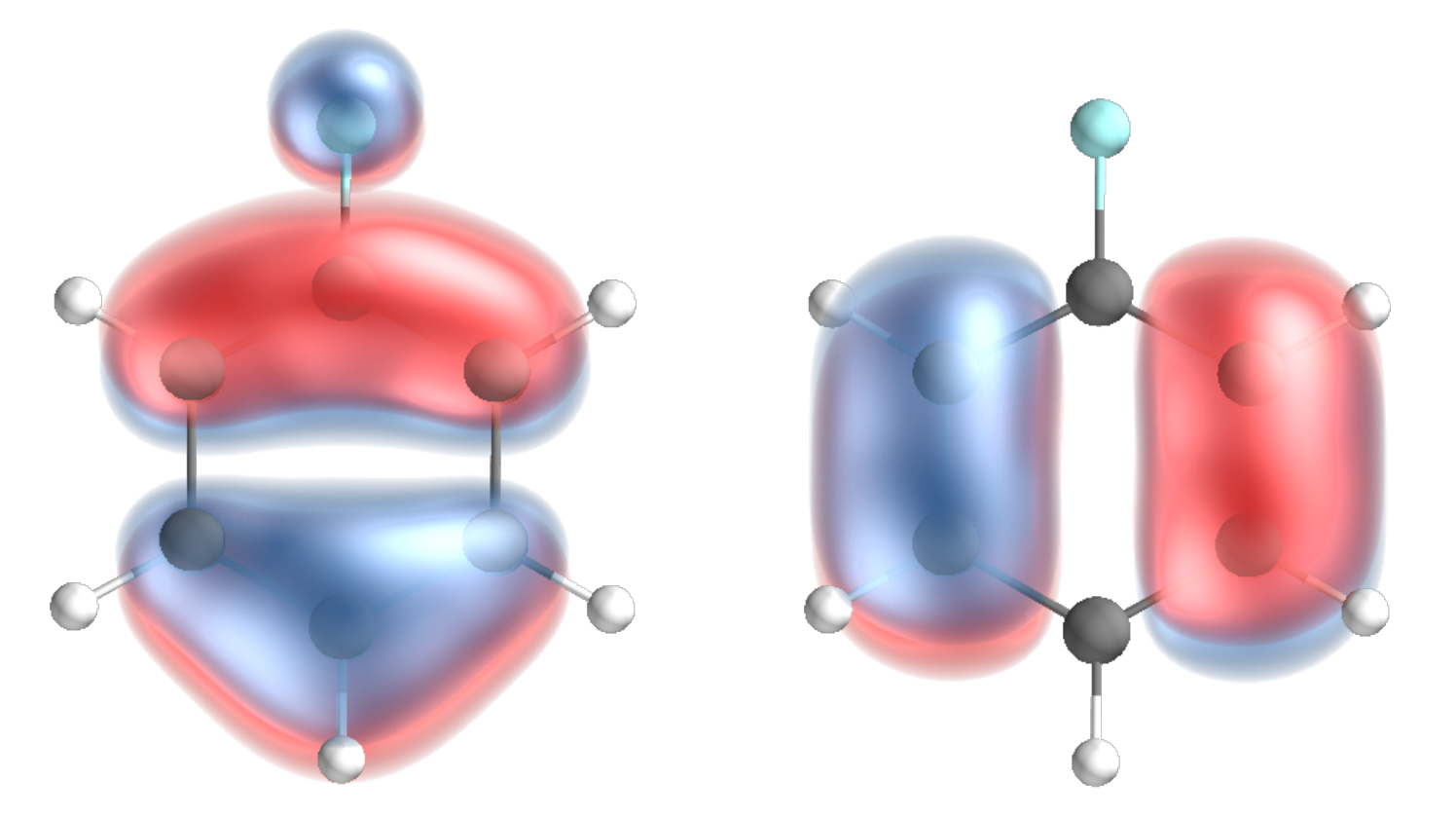}
\caption{Singly occupied molecular orbitals (SOMO) of the (left) $\Psi_Q$ and (right) $\Psi_A$ cationic states of FBZ.}
\label{somo}

\end{figure}

In this letter, the coupled electron-nuclear dynamics upon ionization and excitation of electronic wavepackets in benzene derivatives (Figure\,\ref{somo}) is simulated fully quantum mechanically and in full dimensionality. We focus on ionization which can occur using attosecond pulses (mainly produced in the XUV domain) or intense IR fields. Beyond validating the concept of charge-directed reactivity upon population of different superpositions of the two lowest cationic states of fluoro-benzene, the calculations show that several types of quantum interferences leave clear signatures in the autocorrelation function. Importantly, this quantity is accessible experimentally \textit{via}~HHS~\cite{Austin-2021}.

Benzene (BZ) and its derivatives are a formidable set of systems to investigate coupled electron-nuclear dynamics induced upon ionization by attosecond/sub-femtosecond pulses~\cite{Gindensperger-2007,Vacher-2015-JCP,Despre-2015,Tran-2021}, before electronic decoherence~\cite{Vacher-2015-PRA,Vacher-2017-PRL,Arnold-2017,Despre-2018,Arnold2020,Golubev2020,Matselyukh-2022,DeyKulWor-PRL-22}. Indeed, ionization from the $\pi$ system of BZ 
leads to a conical intersection between the two lowest cationic states. The corresponding two resulting singly occupied orbitals are shown in Figure~\ref{somo} and the  electronic states are said to be of \textit{quinoid} ($\Psi_Q$) and \textit{anti-quinoid}~($\Psi_A$) characters, respectively. In fluoro-benzene (FBZ), a slight offset between the neutral structure and the electronic degeneracy point in the cation leads to a finite energy gap at the vertical ionization geometry:~\mbox{$\Delta E_{\mathrm{vert.}}^{\mathrm{\,adia.}}\approx0.27$~eV}.

In the present simulations of the post-ionization dynamics of BZ and FBZ, several initial electronic wavepackets were considered: the pure diabatic states, $\Psi_Q$ and $\Psi_A$, as well as both equally weighted, real-valued, in-phase and opposite-phase superpositions \textit{i.e.}~\mbox{$\frac{1}{\sqrt{2}}(\Psi_Q\pm\Psi_A)$}. The dynamics was simulated using the single-set formalism of the Direct-Dynamics variational Multi-Configurational Gaussian (DD-vMCG) method, the development of which started two decades ago in the Quantics package~\cite{quantics2020}. In this method, the molecular wavepacket is expanded onto a set of variationally coupled time-dependent Gaussian basis functions (GBF) that evolve quantum mechanically~\cite{Worth-2003,Richings-2015,Vacher-2016-TCA,Tran-2024}. It combines the advantages of quantum dynamics accuracy and \textit{on-the-fly} simulations in full dimensionality. The post-ionization dynamics of BZ and FBZ molecules were simulated for 10~fs using 5 and 10 GBF respectively, to achieve convergence (see SM). 
Electronic structure calculations were performed at the CASSCF(5e, 6o) level \textit{via} the widely used Gaussian software~\cite{Gaussian16} with an active space including the $\pi/\pi^*$ orbitals and using the 6-31G* basis set. Electronic state diabatization was performed using the regularization method~\cite{ThiKop-JCP-99,KopGroMah-JCP-01}. It is noted that the present simulations do not include any external field.

\begin{figure}[t]

\includegraphics[scale=0.21]{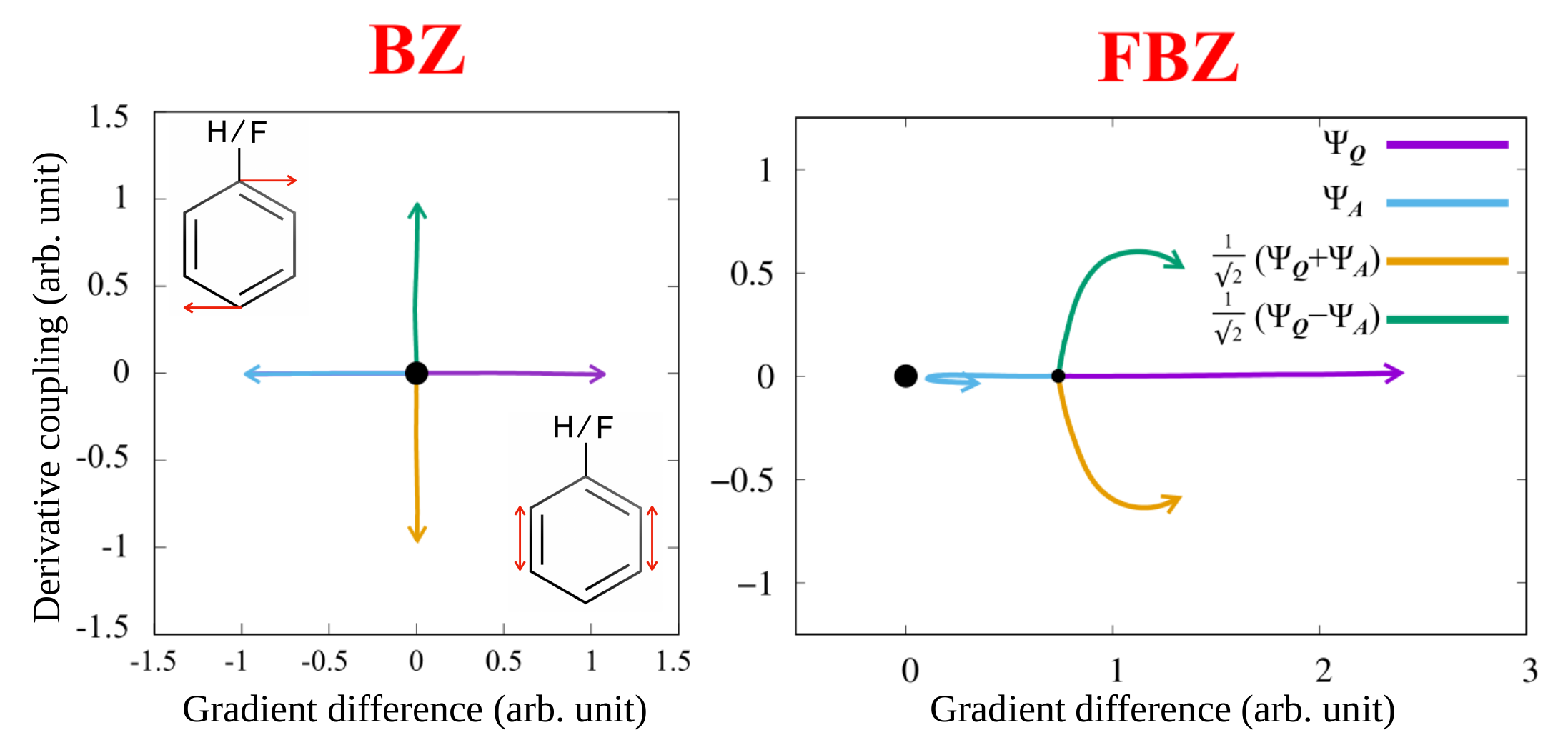}
\caption{Average quantum nuclear motion over 10 fs analysed in the branching space, upon ionization of~(left) BZ and~(right) FBZ molecules and initial population of different electronic wavepackets. The large black circle indicates the conical intersection point while the small black circle marks the vertical ionization geometry (superimposed in BZ). The branching space vectors are represented in the left figure.}
\label{motion_BS}

\end{figure}

Figure~\ref{motion_BS} shows the resulting average nuclear motion \textit{i.e.}~the expectation value of the nuclear position analysed in the branching space of BZ and FBZ, induced by the different wavepackets considered. For both molecules, exciting a pure diabatic state leads to an initial motion oriented solely along the gradient difference vector with opposite directions, while excitation of mixed superpositions, \mbox{$\frac{1}{\sqrt{2}}(\Psi_Q\pm\Psi_A)$}, leads to an initial average motion mostly oriented along the derivative coupling vector, again with opposite directions. 
These observations are in line with previous simulations~\cite{Vacher-2015-JPCA,Meisner-2015,Arnold-2018,Tran-2024}, validating the control over the molecular motion in the branching space achieved by tuning the initial electronic wavepacket composition.

\begin{figure}[t]
\centering
\includegraphics[width=0.44\textwidth]{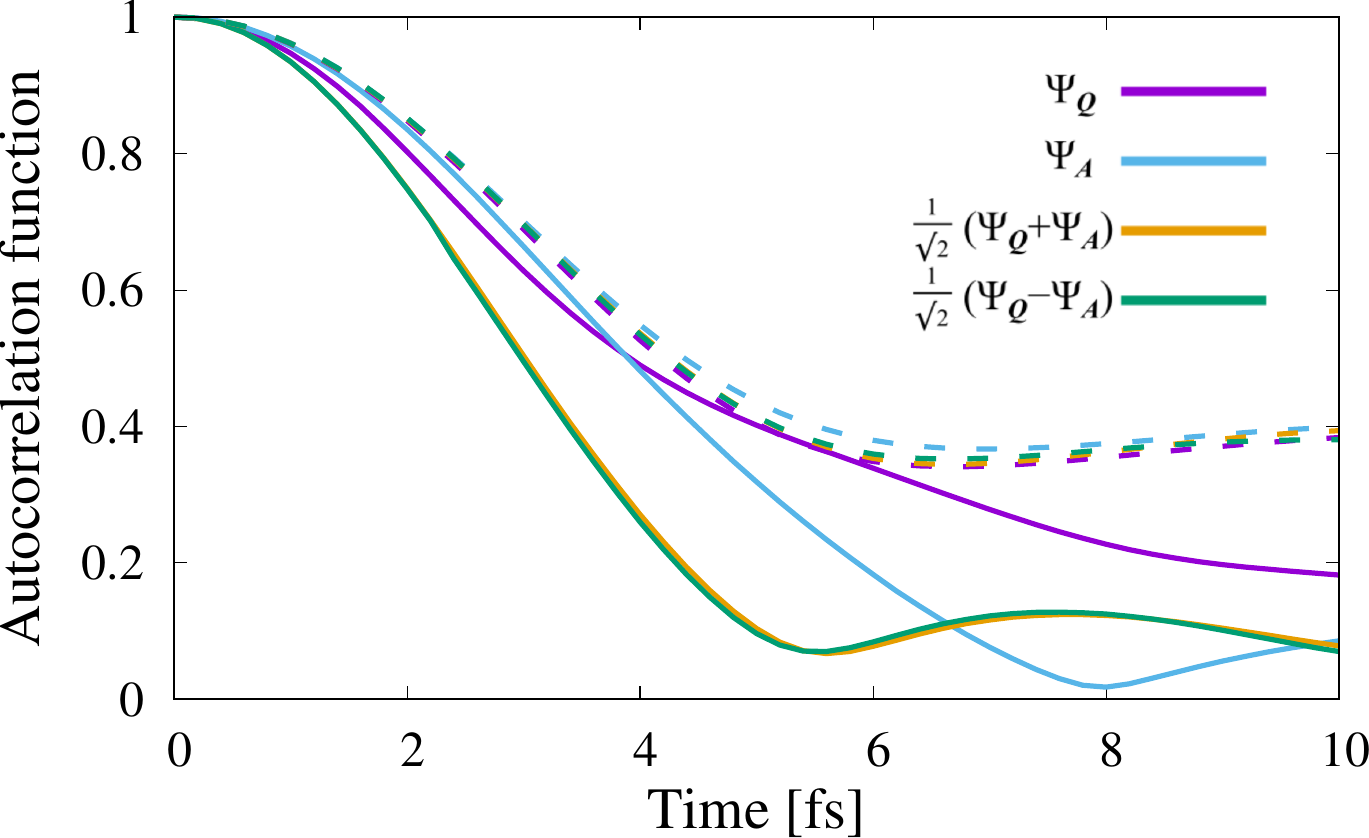}
\caption{Modulus of the autocorrelation functions along the quantum dynamics induced upon ionization of BZ (dashed) and FBZ (solid) and excitation to different electronic wavepackets.}
\label{auto_bz}
\end{figure}

The aims of the present work are (i) to provide an experimental signature of such charge-directed dynamics and (ii) to give physical insights into this processus, using the comparison between BZ and FBZ.
More precisely, we explore features of the autocorrelation function, 
\beq
\mathcal{O}(t) = \Braket{\Psi^\mathrm{mol}(0)|\Psi^\mathrm{mol}(t)},
\eeq
the overlap between the molecular wavepackets $\Psi^\mathrm{mol}$ at time $t$ and at $t=0$ \textit{i.e.}~the ionization time.
Figure \ref{auto_bz} reports the autocorrelation functions for both molecules, upon excitation to the different electronic wavepackets considered. The initial evolution of these autocorrelation functions was fitted using a Gaussian expression, $\exp(-(t/\tau)^2/2)$. The decay times, $\tau$, are reported in Table \ref{table_decay_time}.

\begin{table}[h!]
\centering

\renewcommand{\arraystretch}{1.32}
\setlength{\tabcolsep}{0.1cm}
\caption{Decay times $\tau$ deduced from the fits of the early time evolution of the autocorrelation functions, upon ionization of BZ and FBZ and initial population of different  electronic wavepackets.}
\begin{tabular}{ c c c c c } 
        \hline  
~ & $\ \ \Psi_Q\ \ $  & $\ \ \Psi_A\ \ $ & $\frac{1}{\sqrt{2}}(\Psi_Q+\Psi_A)$ & $\frac{1}{\sqrt{2}}(\Psi_Q-\Psi_A)$\\
\hphantom{F}BZ\ \ & 3.51 fs & 3.56 fs & 3.54 fs & 3.55 fs  \\
FBZ\ \ & 3.05 fs & 3.33 fs & 2.64 fs & 2.64 fs  \\
\hline
\end{tabular}

\label{table_decay_time}
\end{table}

In the BZ cation, both adiabatic states are degenerate at the neutral geometry due to symmetry. Any superposition of the two electronic states being a valid eigenstate,  
BZ is always excited to a pure adiabatic eigenstate. 
All four initial electronic wavepackets yield very similar autocorrelation functions (Figure~\ref{auto_bz}, dashed curves): it steadily decreases from 1 with a decay time of $\approx 3.5$~fs and reaches a minimum value of about 0.4 around 6~fs. 
The predicted decay times are significantly longer than the \mbox{sub-fs} value reported in another theoretical work~\cite{Patchkovskii-2014}. 
The absence of significant differences reflects the isotropy of the potential energy surfaces around the symmetry-required $\Psi_Q/\Psi_A$ conical intersection. 
 
In contrast, in FBZ, the two adiabatic states are not degenerate at the vertical ionization geometry and thus a non-stationary coherent wavepacket can be excited. 
All electronic wavepackets yield autocorrelation functions that drop to significantly lower values (solid curves), almost reaching 0 upon excitation to 
$\Psi_A$.  
Except for the two equally weighted superpositions that are indistinguishable due to the symmetry of the derivative coupling, the initially excited electronic wavepackets lead each to qualitatively different autocorrelation functions. In particular, a significantly faster decay is obtained in the case of the initially mixed electronic wavepackets: $2.6$~fs instead of $3.0-3.3$~fs. 

\begin{figure}[t]
\centering
\includegraphics[width=0.48\textwidth]{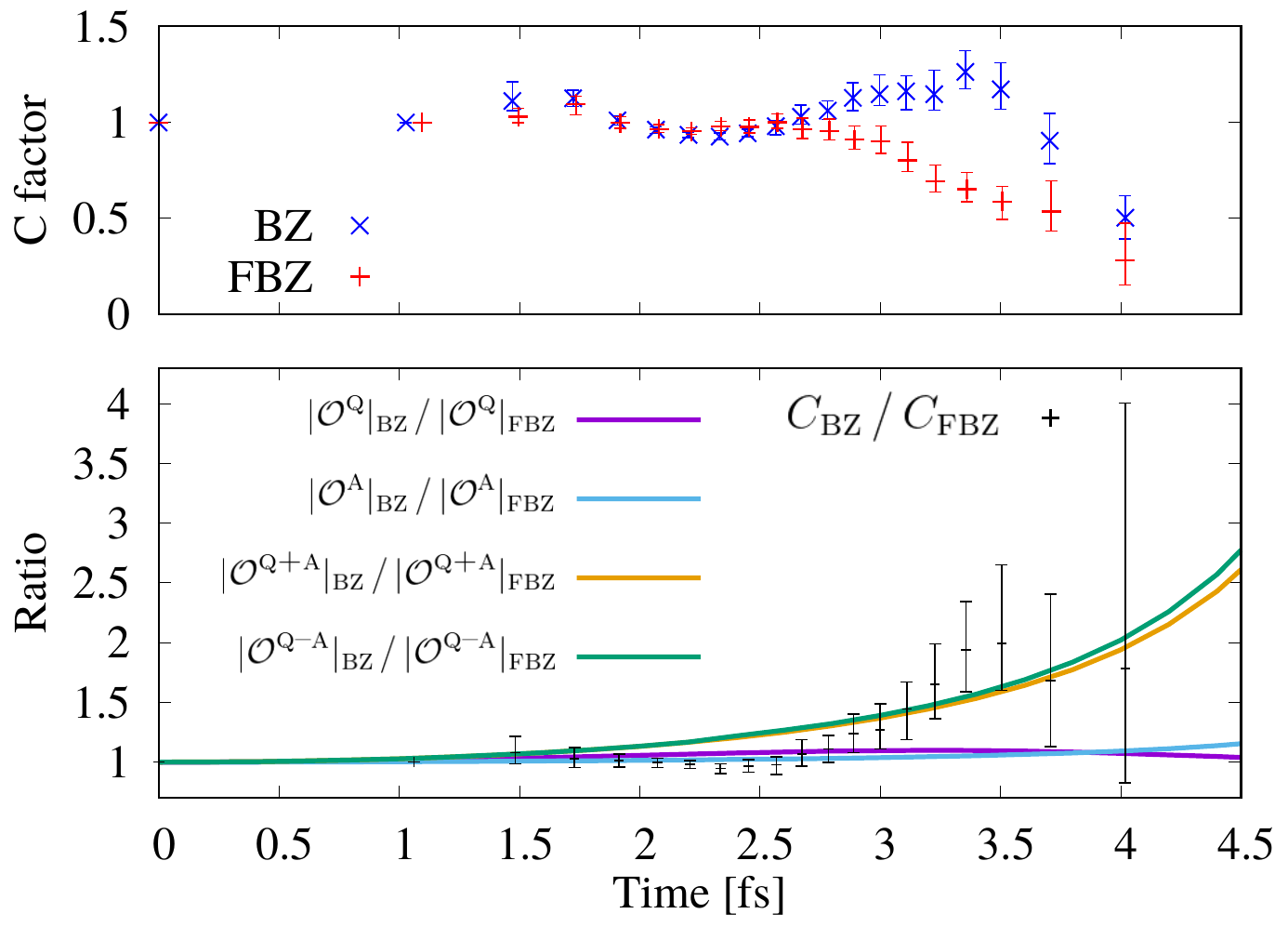}
\caption{(top) Dynamical factors $C$ extracted from HHS of BZ and FBZ. (bottom) Ratio of experimental $C$ factors (black points) and of predicted autocorrelation functions upon ionization and excitation to different electronic wavepackets (solid colored lines).}
\label{exp}
\end{figure}

These predictions are compared with measurements of unaligned BZ and FBZ using the methodology reported previously~\cite{Austin-2021}. In brief, 
2D datasets of harmonic spectrum versus laser intensity were recorded. A physically transparent analysis, in the spirit of quantitative rescattering theory \cite{Lin-2010}, was achieved by fitting the yield to the three-factor ansatz: $S(\omega, I_L)=A(\omega)B(I_L)C(\bar{\omega})$, where $\omega$ is the harmonic frequency and $\bar{\omega}$ is the frequency normalized to the classical cut-off so that the latter occurs at $\bar{\omega}\approx 3.17$. The static factor $A(\omega)$ encodes the photorecombination dipole averaged over participating orbitals and molecular angles, as well as any uncompensated spectral response of the experimental apparatus. The ionization rate factor $B(I_L)$ encodes the angularly and orbital-averaged tunnel ionization rates. The dynamic factor $C(\bar{\omega})$ depends on the normalized frequency which has a one-to-one mapping to the trajectory time, provided short or long trajectories are selected \cite{Mairesse-2003,Lein-2005,Brugnera-2011}. In the present measurements, the short trajectories are selected. The phase matching condition in high harmonic generation strongly favors the electron first returns \cite{Baker-2006}.
This fitting technique implicitly takes into account the different cation potentials in the molecular factors $A$ and $B$ and reduces all dynamic processes, such as the nuclear motion and the electronic dynamics, into a single observable which can be expressed as a function of time directly $C(t)$.
This dynamical factor was shown to be proportional to the nuclear autocorrelation function (the generalized version of which includes the possibility of multiple cationic states) \cite{Lein-2005,Baker-2006,Austin-2021}.

Figure~\ref{exp} (top) presents the experimental $C$ factors for BZ and FBZ. 
By extracting $C$, a proportional measure of the nuclear autocorrelation function is obtained.
From this data, a decay constant of $\tau=4\pm1$~fs was previously extracted for BZ~\cite{Austin-2021}, in excellent agreement with the present calculation. At later times, $C$ falls more rapidly for FBZ. 
By determining the ratio of $C$ between two molecules, the ratio of autocorrelation functions is obtained, free from the influence of the various differing static contributions which can otherwise obscure HHS.
This is possible thanks to the cancellation of the returning electron flux~\cite{Itatani-2004}.
Figure 4 (bottom) plots the ratio of the $C$ factors for BZ and FBZ together with the ratio of predicted autocorrelation functions: good agreement is found for the initial coherent excitation of the two cationic states (see also SM).
Part of the discrepancy between theory and experiment may be explained by the absence of the field in the simulations.

To gain insights into the charge-directed dynamics process and in particular understand further the faster decay of the autocorrelation function in the case of mixed electronic wavepackets, we analyse in more details $\mathcal{O}^\tsc{q+a}(t)$. The system being in a superposition of the two diabatic states, the autocorrelation function is the coherent sum of two complex terms,
\begin{align}
\mathcal{O}^\tsc{q+a}(t) &= \overbrace{\sum_{i,j} {C^\tsc{q}_i}^*(0)\, C^\tsc{q}_j(t) \times \Braket{\Chi_i(0)|\Chi_j(t)}}^{\mathcal{O}^\tsc{q+a}_\tsc{q}(t)} \nonumber\\
& + \underbrace{ \sum_{i,j} {C^\tsc{a}_i}^*(0) \, C^\tsc{a}_j(t) \times  \Braket{\Chi_i(0)|\Chi_j(t)}}_{\mathcal{O}^\tsc{q+a}_\tsc{a}(t)},
\label{auto_corel_eq}
\end{align}
where $\Chi_i$ are the GBF and $C^\tsc{q/a}_i$ the expansion coefficients of the molecular wavepacket in the diabatic basis ``Q/A" and GBF $\Chi_i$. 

\begin{figure}[t]
\centering
\includegraphics[scale=0.33]{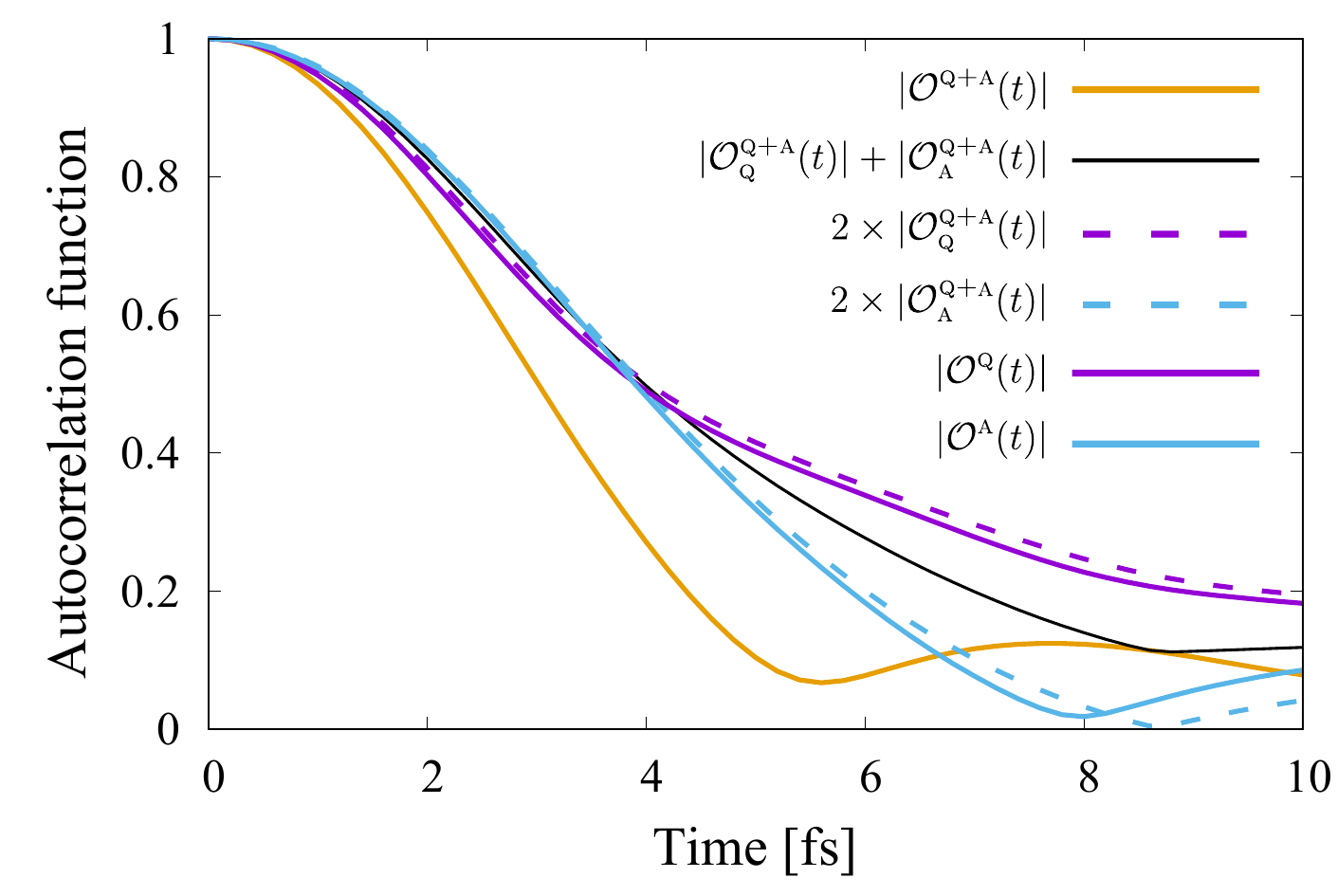}
\caption{Modulus of $\mathcal{O}^\tsc{q+a}(t)$ the autocorrelation function of FBZ upon initial population of the $\frac{1}{\sqrt{2}}(\Psi_Q+\Psi_A)$ wavepacket (orange) and of its two diabatic components $\mathcal{O}^\tsc{q+a}_\tsc{q}(t)$~(dashed~purple) and  $\mathcal{O}^\tsc{q+a}_\tsc{a}(t)$~(dashed~blue) -- note that they were doubled here for ease of representation. Incoherent sum $|\mathcal{O}^\tsc{q+a}_\tsc{q}(t)|+|\mathcal{O}^\tsc{q+a}_\tsc{a}(t)|$ (black). Modulus of the autocorrelation functions upon initial population of the pure diabatic states, $\mathcal{O}^\tsc{q}(t)$~(purple) and $\mathcal{O}^\tsc{a}(t)$ (blue).}
\label{interference_auto}
\end{figure}

Figure \ref{interference_auto} decomposes the full autocorrelation function $\mathcal{O}^\tsc{q+a}(t)$ (orange), onto its two diabatic components $\mathcal{O}^\tsc{q+a}_\tsc{q}(t)$ (dashed purple)~and  $\mathcal{O}^\tsc{q+a}_\tsc{a}(t)$~(dashed blue). Is also reported the incoherent sum \mbox{$|\mathcal{O}^\tsc{q+a}_\tsc{q}(t)|+|\mathcal{O}^\tsc{q+a}_\tsc{a}(t)|$ (black)}. 
The naive incoherent summation of both diabatic components (black) yields a result very different from the true autocorrelation function (orange): slower decay with no structure. This demonstrates that interferences between the two diabatic components are responsible for the accelerated decay of the autocorrelation function. 
This effect is named \textit{inter-state interference} hereafter. 
It is noted that the autocorrelation functions of the diabatic components are very similar to those of the pure diabatic states. In SM we use this to model the autocorrelation functions of arbitrary initial wavepackets. 
 
For both $\mathcal{O}^\tsc{q+a}_\tsc{q/a}(t)$ terms to interfere, a phase difference has to accumulate along the dynamics. 
Importantly, this phase difference is independent of the initial phases of the two diabatic components. 
It only accounts for the phase difference accumulated during the dynamics 
which, over a short period of $\tau$ a.u., is approximately $\tau \times \Delta E_{\mathrm{vert.}}^{\mathrm{\,adia.}}$.  
In FBZ, the phase difference initially accumulates at the expected rate given the initial energy gap (see~SM). 
Notably, a phase difference of $\pi$, that maximizes interference effects, is reached just before 6~fs which corresponds to the time at which the full autocorrelation function (Figure~\ref{interference_auto}, orange curve) presents a minimum. In the case of BZ, the two electronic states being degenerate at the vertical ionization geometry, the phase difference 
accumulated throughout the dynamics remains extremely small.

After the inter-state interference local minimum of $|\mathcal{O}^\tsc{q+a}(t)|$ in FBZ, a small increase can be observed. This originates from the antiquinoid diabatic component of the autocorrelation function $\mathcal{O}^\tsc{q+a}_\tsc{a}(t)$ (dashed blue) almost reaching zero around the 8.5 fs mark, thus suppressing the inter-state interference.

In the following, the important decay of the full autocorrelation function upon excitation to the pure $\Psi_A$ diabatic state $\mathcal{O}^\tsc{a}(t)$ -- a feature also present in $\mathcal{O}^\tsc{q+a}_\tsc{a}(t)$ -- is analysed in detail.
One of the reasons for it is 
the large diabatic population transfer: final quinoid population $\approx 0.46$. This also explains the direction reversal of the average nuclear motion along the gradient difference vector (Fig.\,\ref{motion_BS}, right). However, this alone does not explain the full structure of $\mathcal{O}^\tsc{a}(t)$. 

\begin{figure}[t]
\includegraphics[scale=0.32, angle=-90]{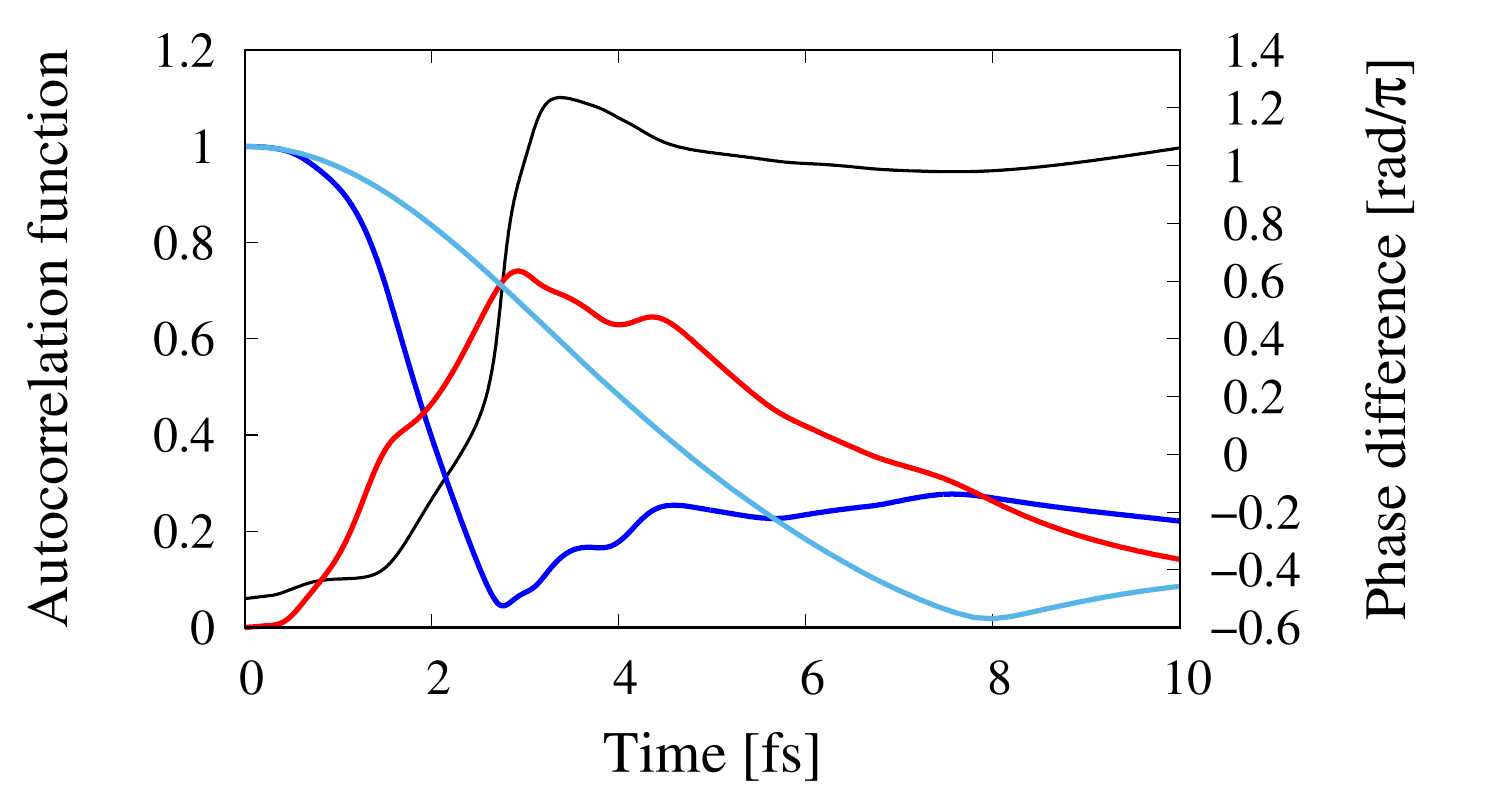}
\caption{Modulus of $\mathcal{O}^\tsc{a}(t)$ the autocorrelation function of the FBZ cation upon initial population of the pure $\Psi_A$ diabatic state (light blue) and of its two space-restricted components $\mathcal{O}^\tsc{a}_+(t)$~(blue) and  $\mathcal{O}^\tsc{a}_-(t)$~(red). Phase difference between the two space-restricted components of the autocorrelation function (black).}
\label{intra_state_interference}
\end{figure}

Figure \ref{intra_state_interference} reports the autocorrelation function of FBZ upon initial population of the pure $\Psi_A$ diabatic state (light blue) and of its spatial components on either side of the conical intersection seam. More precisely, it is decomposed as 
\beq
\mathcal{O}^\tsc{a}(t)=\mathcal{O}^\tsc{a}_+(t) + \mathcal{O}^\tsc{a}_-(t).
\eeq
where $\mathcal{O}^\tsc{a}_+(t)$ and $\mathcal{O}^\tsc{a}_-(t)$ are computed by only accounting for GBF whose final position are either, on the same side of the conical intersection seam as the vertical ionization geometry \textit{i.e.}~\mbox{$E^{\mathrm{diab.}}_A-E^{\mathrm{diab.}}_Q>0$~($\mathcal{O}^\tsc{a}_+$,~blue)}, or on the opposite side:~$E^{\mathrm{diab.}}_A-E^{\mathrm{diab.}}_Q<0$~($\mathcal{O}^\tsc{a}_-$,~red) (see SM).
Around the 8 fs mark, both components have equal modulus but a phase difference (black) of about $\pi$ radian: the near cancellation of $\mathcal{O}^\tsc{a}(t)$ stems from destructive interference between the space-restricted components $\mathcal{O}^\tsc{a}_+(t)$ and  $\mathcal{O}^\tsc{a}_-(t)$ belonging to the same $\Psi_A$ diabatic state. 
This effect is named \textit{intra-state interference}.

In summary, we report quantum simulations of the dynamics induced in the BZ and FBZ molecules upon ionization and excitation to different electronic wavepackets. In BZ, all considered initial electronic states lead to similar dynamics due to the isotropy of the potential energy surfaces.
In stark contrast, the dynamics in FBZ 
exhibit different types of interference effects. 
In particular, inter-state interference between the two diabatic components leads to an accelerated decay of the autocorrelation function upon excitation of a mixed electronic wavepacket. Also, an intra-state interference leads to the almost cancellation of the autocorrelation function upon population of the pure higher lying diabatic state. This is due to the destructive interference between the components 
of the wavepacket that crosses the conical intersection seam during the dynamics and the part that ultimately remains on the same side. 

The predicted autocorrelation functions are in very good agreement with experimentally measured dynamics via HHS in both molecules, highlighting the sensitivity of autocorrelation functions to the composition of the initial electronic wavepacket. Moreover, in the case of FBZ, this provides a clear signature of, not only the ability to influence the induced molecular dynamics through electronic wavepackets, but also of the coherent behavior and interference effects that may influence chemical dynamics on ultrafast timescales.
The present simulations indicate the coherent excitation of several electronic states in the experiment. However, they also show that the current experimental signal to noise ratio does not allow us to draw any decisive conclusions as to the exact composition of the initial wavepacket or the ensemble of initial wavepackets (see also SM). We thus hope that the present study will motivate further experimental investigations of such quantum ultrafast dynamical effects with improved precision, perhaps with aligned sample and controlled polarization to affect the composition of the excited wavepacket. Using a longer wavelength would also extend the experimental time window and allow direct observations of the structures in the autocorrelation functions due to interferences.

\begin{acknowledgements}
We thank the \textit{Région des Pays de la Loire} who provided post-doctoral funding for A.F. The project is also partly funded by the European Union through ERC grant 101040356 (M.V.). Views and opinions expressed are however those of the authors only and do not necessarily reflect those of the European Union or the European Research Council Executive Agency. Neither the European Union nor the granting authority can be held responsible for them. This work was performed using HPC resources from GENCI-IDRIS (Grant 101353) and CCIPL (Le centre de calcul intensif des Pays de la Loire).
\end{acknowledgements}

\bibliographystyle{ieeetr}
\bibliography{references}

\end{document}


\title{Supplemental Material: Signature of attochemical quantum interference upon ionization and excitation of an electronic wavepacket in fluoro-benzene}

\author{Anthony Fert\'e}
\affiliation{Nantes Universit\'e, CNRS, CEISAM, UMR 6230, F-44000 Nantes, France}
\author{Dane Austin}
\author{Allan S. Johnson}
\author{Felicity McGrath}
\affiliation{Quantum Optics and Laser Science Group, Blackett Laboratory, Imperial College London, London, UK}
\author{Jo\~ao Pedro Malhado}
\affiliation{Chemistry Department, Imperial College London, Prince Consort Road, London, SW7 2AZ, UK}
\author{Jon P. Marangos}
\affiliation{Quantum Optics and Laser Science Group, Blackett Laboratory, Imperial College London, London, UK}
\author{Morgane Vacher}\email{morgane.vacher@univ-nantes.fr}
\affiliation{Nantes Universit\'e, CNRS, CEISAM, UMR 6230, F-44000 Nantes, France}

\date{\today}

\maketitle

%


\section*{Convergence of the DD-vMCG dynamics}
\begin{figure}[h]
\includegraphics[width=0.4\textwidth, angle=-90]{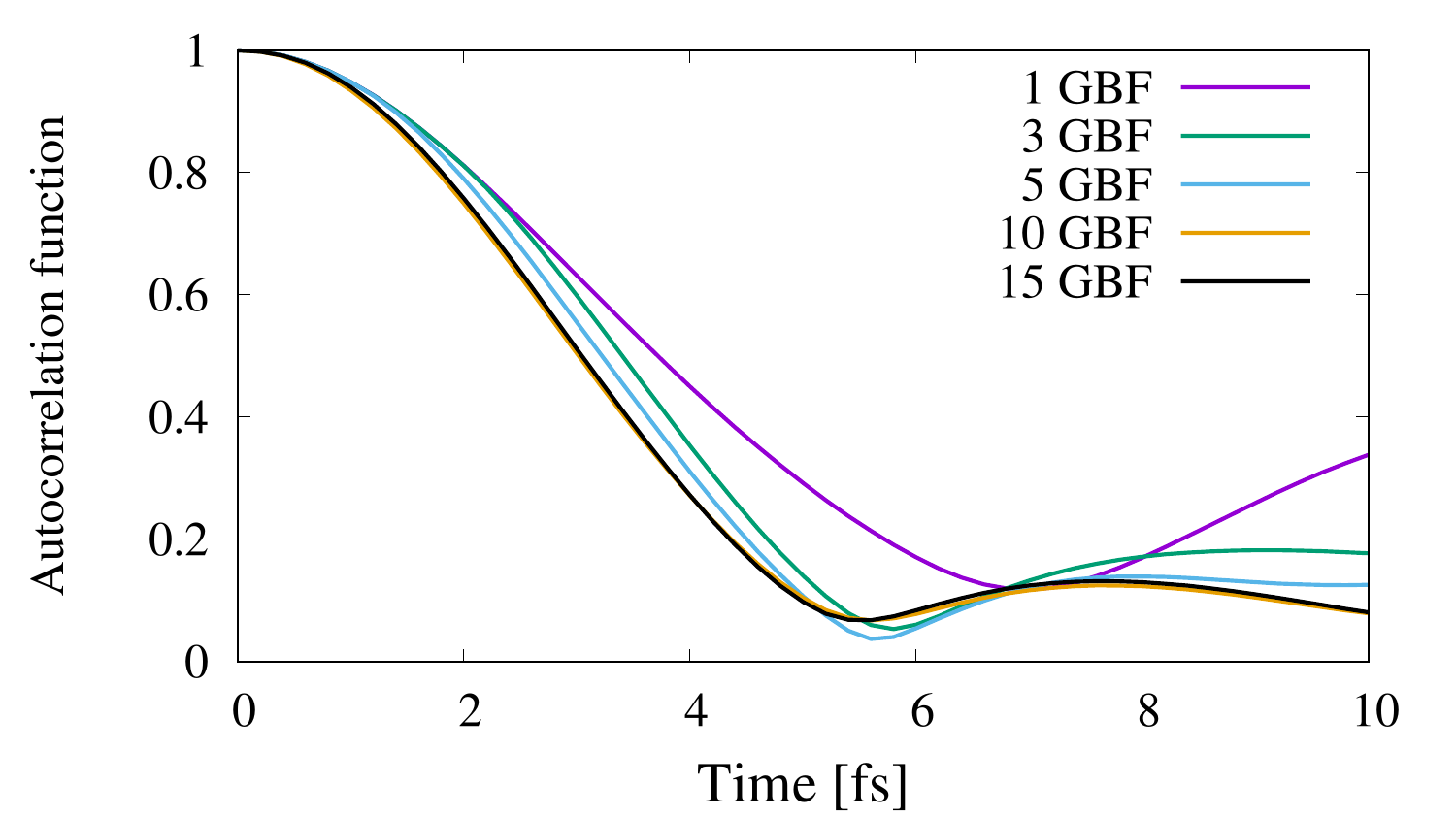}
\caption{Modulus of $\mathcal{O}^\tsc{q+a}(t)$ the autocorrelation function of FBZ upon initial population of the $\frac{1}{\sqrt{2}}(\Psi_Q+\Psi_A)$ cationic wavepacket during the DD-vMCG dynamics simulation: 1 GBF (purple), 3 GBF (green), 5~GBF~(blue), 10 GBF (orange), and 15 GBF (black).}
\label{convergence_auto}
\end{figure}

In order to asses the convergence of the DD-vMCG simulations, the number of GBF is increased until stabilization of the autocorelation functions. Figure \ref{convergence_auto} reports, as an example, the autocorrelation function computed along the dynamics of the FBZ cation upon initial population of the $\frac{1}{\sqrt{2}}(\Psi_Q+\Psi_A)$ wavepacket and for different number of GBF. As one can see, both orange and black curves, respectively obtained with 10 and 15 GBF, are superimposed.

\section*{BZ and deuterated BZ dynamics}

\begin{figure}[h]
\includegraphics[scale=0.34]{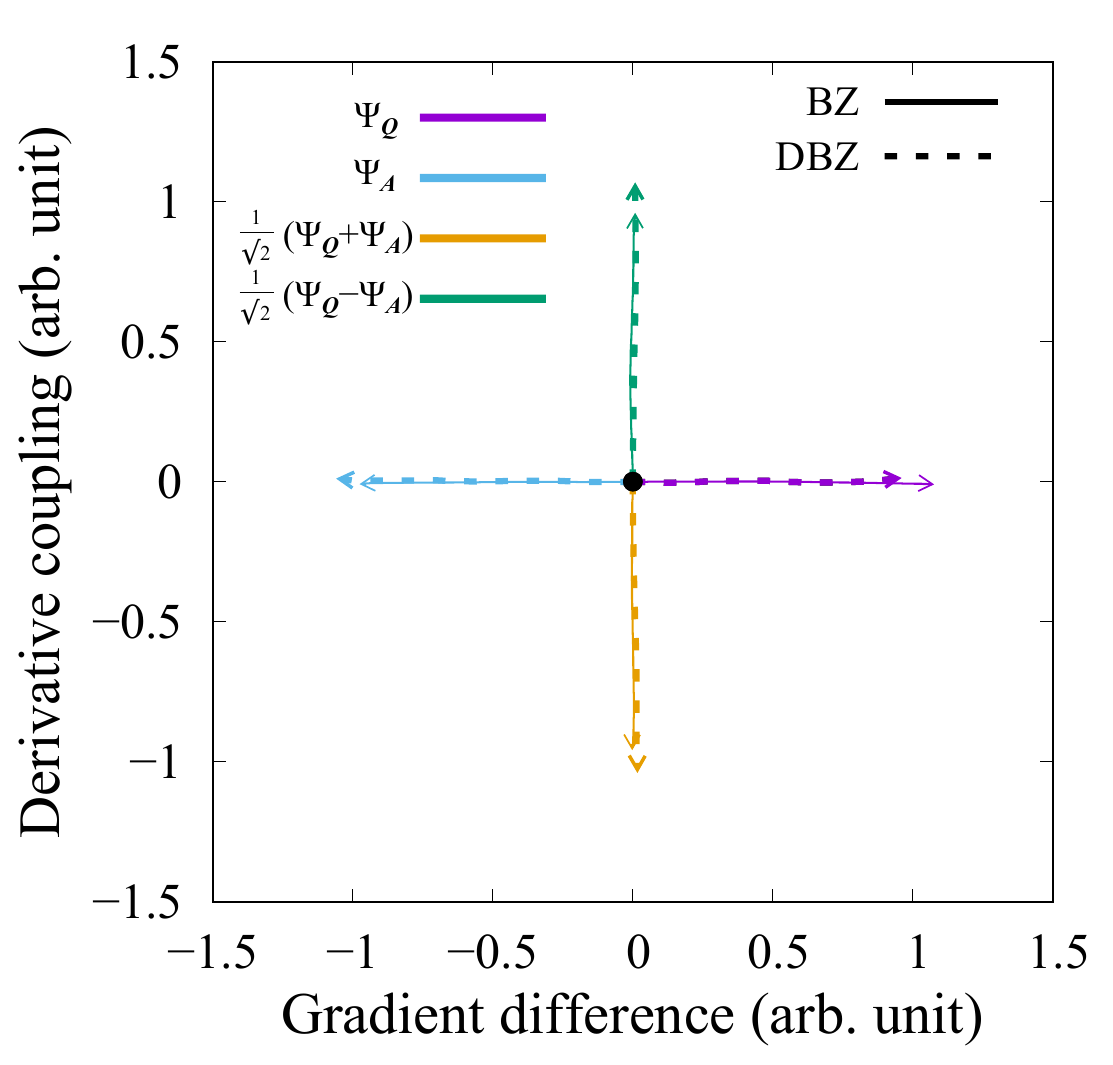}\hspace{+1cm}
\includegraphics[scale=0.33]{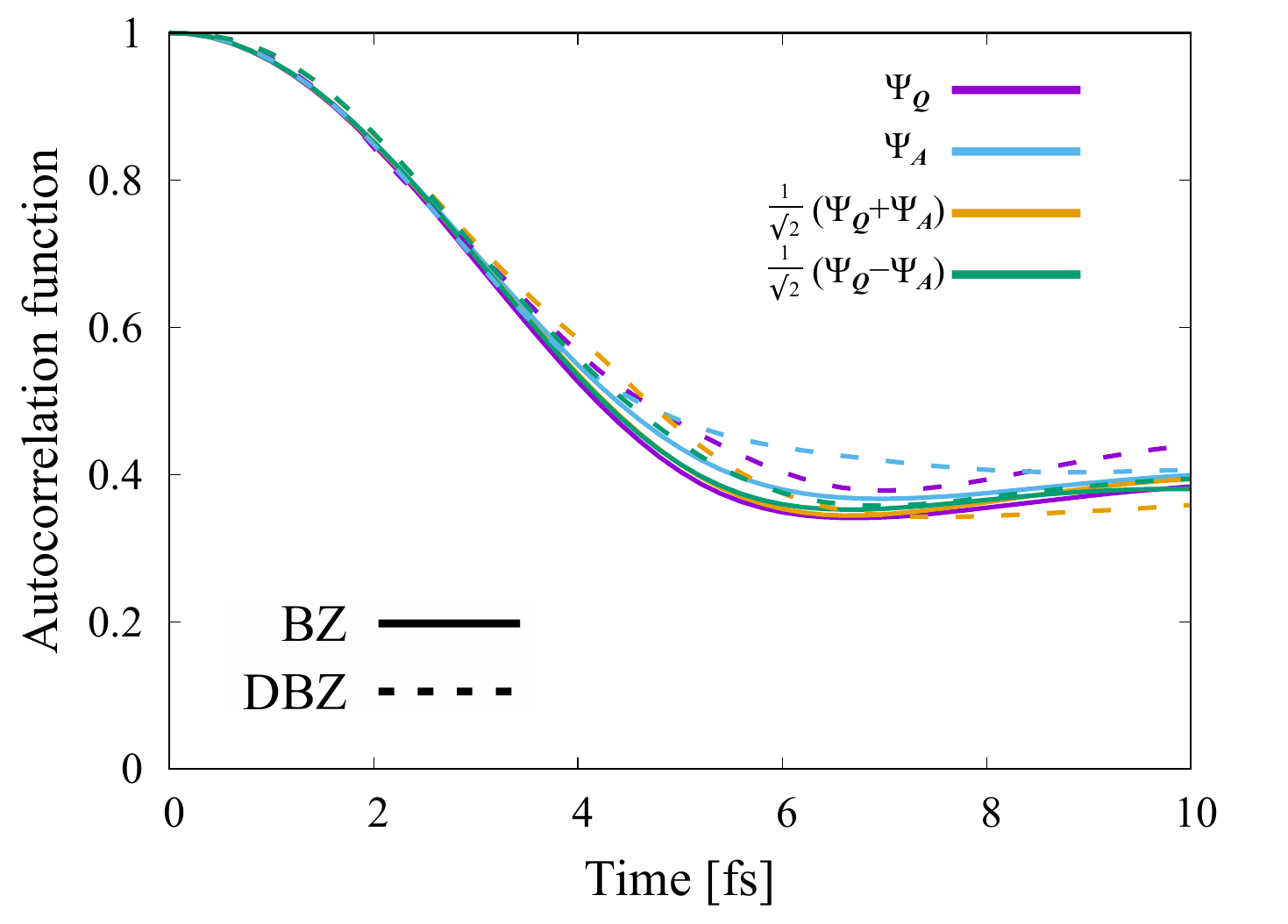}
\caption{Left: Average quantum nuclear motion in the branching space over 10 fs and upon ionization of BZ~(plain) and DBZ~(dashed) cation upon initial population of $\Psi_Q$~(purple), $\Psi_A$~(blue), \mbox{$\frac{1}{\sqrt{2}}(\Psi_Q+\Psi_A)$~(orange)} and \mbox{$\frac{1}{\sqrt{2}}(\Psi_Q-\Psi_A)$~(green)}. Right: Modulus of the autocorrelation functions along the quantum dynamics induced upon ionization of BZ (plain) and DBZ (dashed) molecules and excitation to $\Psi_Q$~(purple), $\Psi_A$~(blue), $\frac{1}{\sqrt{2}}(\Psi_Q+\Psi_A)$~(orange) and \mbox{$\frac{1}{\sqrt{2}}(\Psi_Q-\Psi_A)$~(green)} electronic wavepackets.}
\label{DBZ_SI}
\end{figure}

Figure \ref{DBZ_SI} compares the dynamics of BZ (plain) and deuterated benzene~(DBZ, dashed) cations upon population of $\Psi_Q$~(purple), $\Psi_A$~(blue), \mbox{$\frac{1}{\sqrt{2}}(\Psi_Q+\Psi_A)$~(orange)} and \mbox{$\frac{1}{\sqrt{2}}(\Psi_Q-\Psi_A)$~(green)} wavepackets. The BZ (plain curves) and deuterated BZ (dashed curves) display highly similar dynamics and autocorrelation functions. This is explained by the fact that the induced nuclear dynamics does not significantly involve the carbon-hydrogen or carbon-deuterium bonds; it almost only involve the carbon ring as only the $\pi$ conjugated system is involved in the ionization and excitation process.

\section*{Diabatic decomposition of the autocorelation function of BZ and FBZ mixed wavepackets}

\begin{figure}[h]
\includegraphics[scale=0.31]{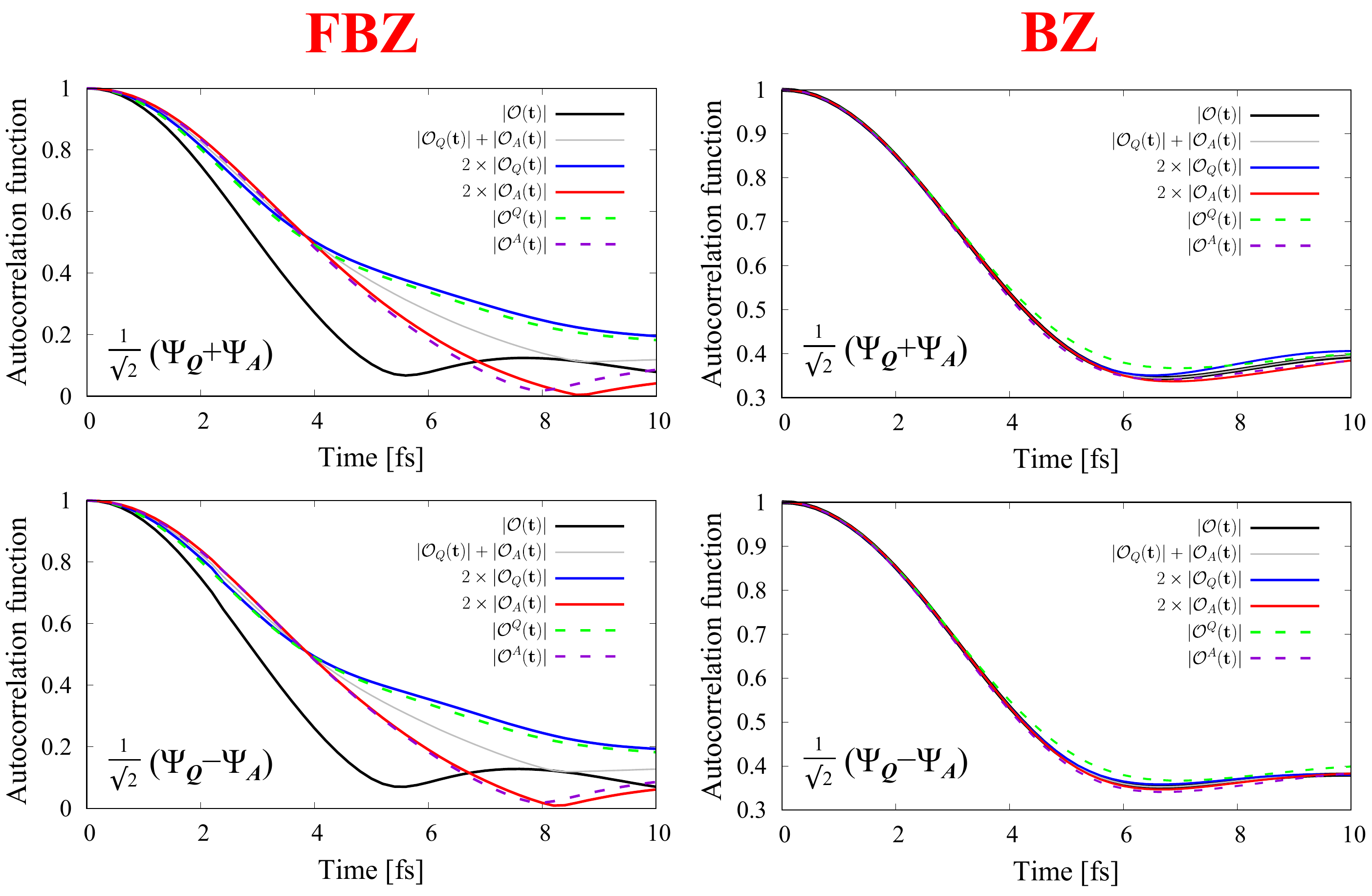}
\caption{Modulus of the autocorrelation functions of the FBZ (left) and BZ (right) cations upon initial population of an equally weighted superposition of the two diabatic states (black). The initial composition of the wavepacket is reported in the bottom left corners of each figures. Diabatic components of the full autocorrelation function, $\mathcal{O}_Q(t)$ (blue) and  $\mathcal{O}_A(t)$ (red), note that they were doubled here for ease of representation. Incoherent sum $|\mathcal{O}_Q(t)|+|\mathcal{O}_A(t)|$ (grey). Modulus of the autocorrelation functions of the cations upon initial population of a pure diabatic state, $\mathcal{O}^Q(t)$ (dashed green) and $\mathcal{O}^A(t)$ (dashed purple).}
\label{interference_auto_SI}
\end{figure}

Figure \ref{interference_auto_SI} shows similar diabatic decompositions of the autocorrelation functions (as reported in the main text of this letter) for both $\frac{1}{\sqrt{2}}(\Psi_Q\pm\Psi_A)$ mixed wavepackets and for both BZ (right) and FBZ~(left) molecules. Clearly, there is no inter-state interference effects in the case of BZ contrary to the FBZ case. Indeed, in the former case, the modulus of the full autocorrelation function (black) is equal to the incoherent sum of the modulus of the two diabatic components (grey). As thoroughly discussed in the main body of this letter, this is not the case for FBZ. Moreover, this figure illustrates the symmetry that exists between both in phase (top row) and opposite phase (bottom) wavepackets due to the symmetry of the system with respect to the molecular displacement along the derivative coupling vector. Indeed, for both systems, despite leading to molecular displacement in opposite direction with respect to the derivative coupling vector, both wavepackets yields indistinguishable autocorrelation functions. 

\section*{Inter-state interference effects and independence with respect to the initial phases of the diabatic components of the wavepacket}

\begin{figure}[h]
\centering
\includegraphics[scale=0.5, angle=-90, trim=  0 0 -0.8cm 0, clip]{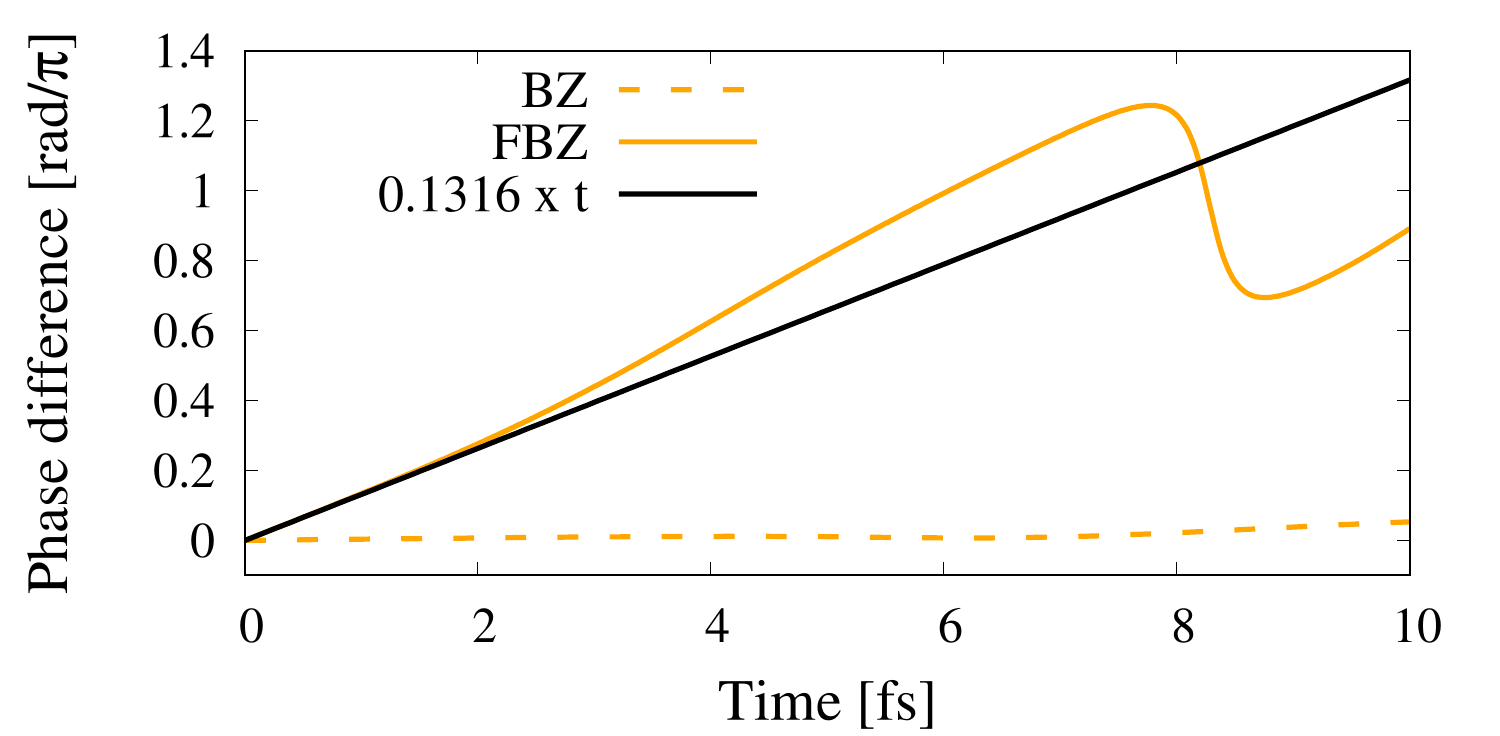}
\caption{Phase difference $\Delta\varphi$ between the diabatic components of the BZ (dashed) and FBZ (plain) autocorrelation functions upon population of $\frac{1}{\sqrt{2}}(\Psi_Q+\Psi_A)$. We also report the $0.1316 \times t$ line (black).}
\label{phase_auto}
\end{figure}

Figure \ref{phase_auto} reports the phase difference, $\Delta \varphi (t)$, after initial excitation of $\frac{1}{\sqrt{2}}(\Psi_Q+\Psi_A)$ in BZ (dashed) and FBZ (plain orange): $\Delta \varphi (t) = \mathrm{arg}\big(\mathcal{O}^\tsc{q+a}_\tsc{q}(t)\big)-\mathrm{arg}\big(\mathcal{O}^\tsc{q+a}_\tsc{a}(t)\big)$, with $\mathrm{arg}(X)$ being the argument of the complex number~$X$.
One easily obtains that 
\beq
| \mathcal{O}^\tsc{q+a}(t) |^2 = | \mathcal{O}^\tsc{q+a}_\tsc{q}(t) |^2 + | \mathcal{O}^\tsc{q+a}_\tsc{a}(t) |^2 
+ | \mathcal{O}^\tsc{q+a}_\tsc{q}(t) | \times | \mathcal{O}^\tsc{q+a}_\tsc{a}(t) | \times 2  \cos\big( \Delta \varphi (t)\big). 
\eeq

Formally, each diabatic component can be expanded in the adiabatic basis. In the present case, near the vertical ionization geometry, both $\Psi_Q$ and $\Psi_A$ are very strongly dominated by a single adiabatic state (more than 90\% of their total norm). Therefore, the phase difference accumulated over a short period of $\tau$ a.u. is approximately $\tau \times \Delta E_{\mathrm{vert.}}^{\mathrm{\,adia.}}$. 
In FBZ, the initial $\approx 0.27$~eV gap should lead to a phase difference that accumulates at a rate of about 0.1316$\times \pi$~rad/fs (black line in Fig.~\ref{phase_auto}).
In the case of BZ, the vertical ionization geometry corresponds to the conical intersection, and thus the rate at which the phase difference is expected to accumulate is null. Obviously, the molecular wavepacket is not fully localised on the conical intersection and spreads and moves with time in non-degenerate regions, explaining the slight phase difference accumulated during the dynamics. However, subsequent interference effects remain fairly negligible.

\begin{figure}[h]
\includegraphics[scale=0.32]{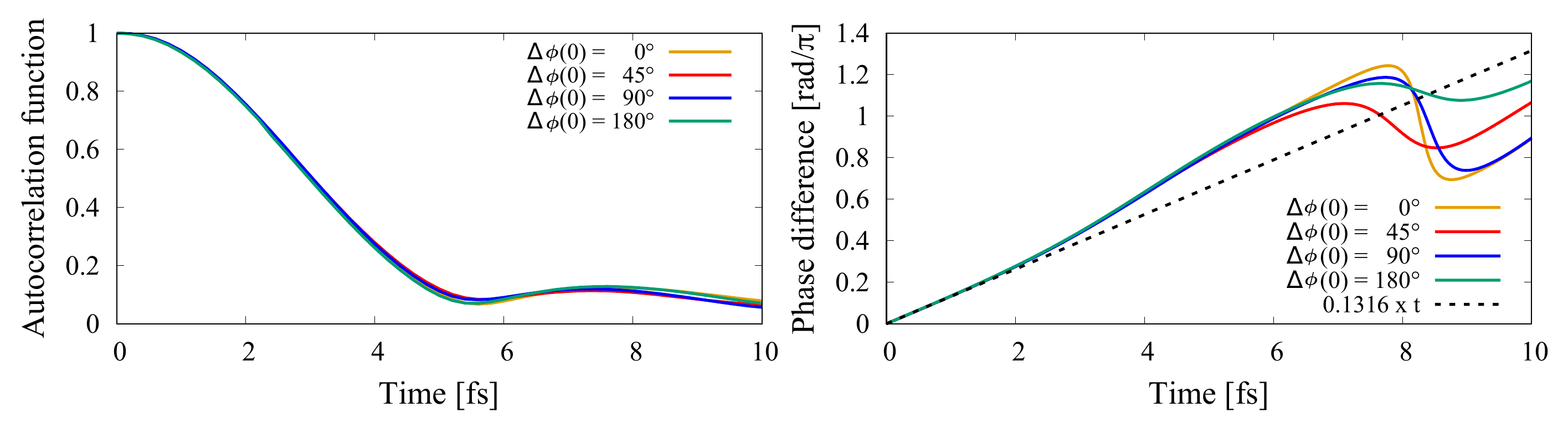}
\caption{Upon population of an equally weighted superposition of the two diabatic states of the FBZ molecule with various initial phases \textit{i.e.} $\frac{1}{\sqrt{2}}(\Psi_Q+\Psi_A \mathrm{e}^{i \Delta \phi})$. Left: Modulus of the autocorrelation function. Right: Accumulated phase difference ($\Delta \varphi$) between the diabatic components of the autocorrelation function. Orange : $\Delta \phi = 0$°. Red~:~$\Delta \phi = 45$°. Blue : $\Delta \phi = 90$°. Green : $\Delta \phi = 180$°.}
\label{init_phase_SI}
\end{figure}

As stated in the main body of this letter, the inter-state interference effects observed in case of a mixed cationic wavepacket of FBZ are independent of the absolute and relative phases of the diabatic components of the initial wavepacket. To illustrate this property, the left panel of figure \ref{init_phase_SI} reports the modulus of the autocorrelation function of the FBZ cation upon initial population of equally weighted wavepacket of the following form, $\frac{1}{\sqrt{2}}(\Psi_Q+\Psi_A \mathrm{e}^{i \Delta \phi})$, for a series of initial electronic phase differences $\Delta\phi$. The right panel of figure \ref{init_phase_SI} also reports the phase difference $\Delta \varphi$ between the two diabatic components of the autocorrelation function. As one can see, modulating the initial relative phases $\Delta\phi$ do not yield any modification of the subsequent autocorrelation function. 

\section*{Phase accumulation rate of the diabatic components of the autocorelation function}

\begin{figure}[h]
\includegraphics[scale=0.45]{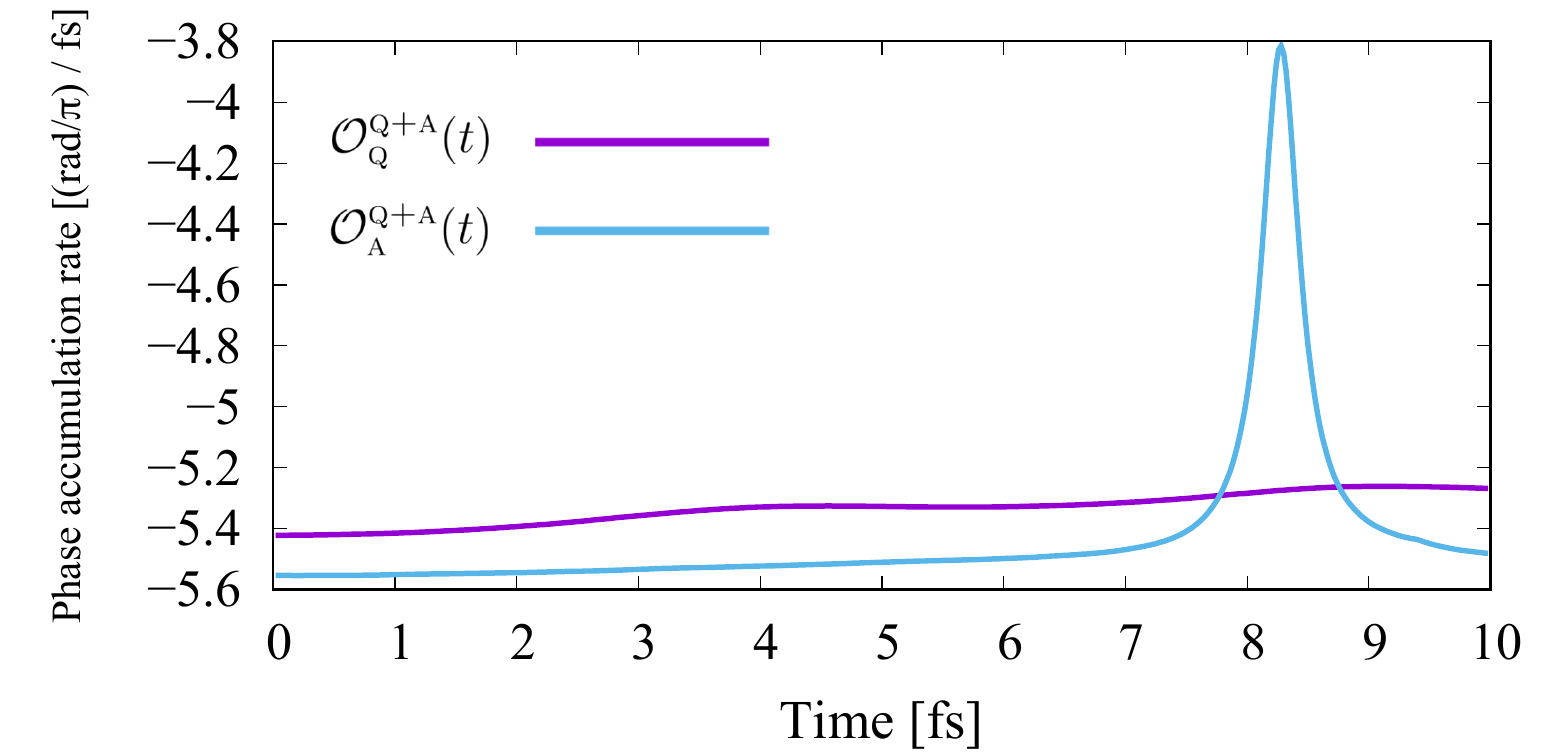}
\caption{Time derivative of the phase of the diabatic components of the full FBZ cation autocorrelation function upon excitation to the $\frac{1}{\sqrt{2}}(\Psi_Q+\Psi_A)$ wavepacket. Purple: Time derivative of the complex phase of $\mathcal{O}^\tsc{q+a}_\tsc{q}(t)$. Blue: Time derivative of the complex phase of $\mathcal{O}^\tsc{q+a}_\tsc{a}(t)$. }
\label{phase_derivative}
\end{figure}

Figure \ref{phase_derivative} reports the derivative with respect to time of the phase of each diabatic components of the full autocorrelation function of the FBZ cation upon initial excitation of the $\frac{1}{\sqrt{2}}(\Psi_Q+\Psi_A)$ wavepacket. This figure clearly illustrates that the phase of the $\mathcal{O}^\tsc{q+a}_\tsc{q}(t)$ term accumulates quasi linearly with time while the $\mathcal{O}^\tsc{q+a}_\tsc{a}(t)$ term presents an important variation of its accumulation rate over a short time period around 7.5 to 9.0 fs. This variation in the time derivative of the phase of $\mathcal{O}^\tsc{q+a}_\tsc{a}(t)$ leads to the step-like feature observed in the evolution of the $\Delta \varphi$ phase difference.

\section*{Spatial partition of the autocorrelation function}

The partitioning of the full autocorrelation function of the FBZ cation upon initial population of the anti-quinoid diabatic state, in the two $\mathcal{O}^\tsc{a}_+(t)$ and $\mathcal{O}^\tsc{a}_-(t)$ space restricted components was obtained by only accounting for the GBF whose final positions are on a given side of the conical intersection seam. This decomposition is motivated by the results displayed in Fig.\,\ref{GBF_traj}~and~\ref{GBF_diab_E_diff}. The former shows the trajectory in the branching plane of each GBF during the dynamics induced upon population of $\Psi_A$. On the other hand, the later shows the diabatic energy gap at the position of each GBF centers. The GBF partitioning used in the decomposition of the autocorrelation function is depicted by the color coding in these figures. 

%

\begin{figure*}
\centering
	\begin{minipage}{.475\linewidth}
	\includegraphics[scale=0.3, angle=-90]{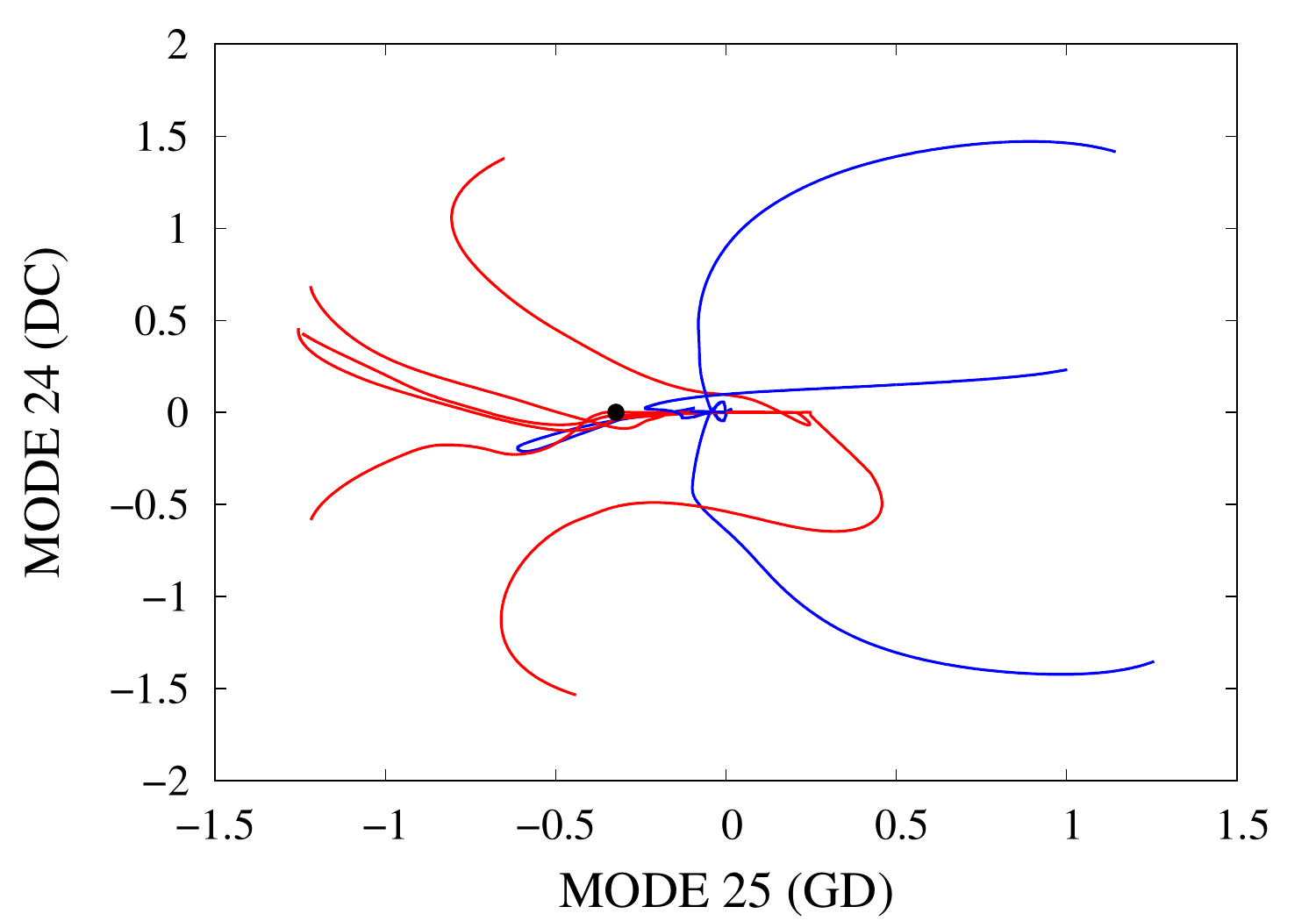}
	\caption{Trajectory, in the plane spanned by the vector of the normal mode 24 (representative of the DC) and 25 (representative of the GD), of the 10 GBF during the DD-vMCG simulation of the non-adiabatic dynamics of the FBZ cation upon initial excitation of the pure $\Psi_A$ diabatic state. Colors are used to highlight similar behavior within the GBF trajectories.}
	\end{minipage}%
	\begin{minipage}{.05\linewidth}
	~
	\end{minipage}%
    \begin{minipage}{.475\linewidth}
    \includegraphics[scale=0.3, angle=-90]{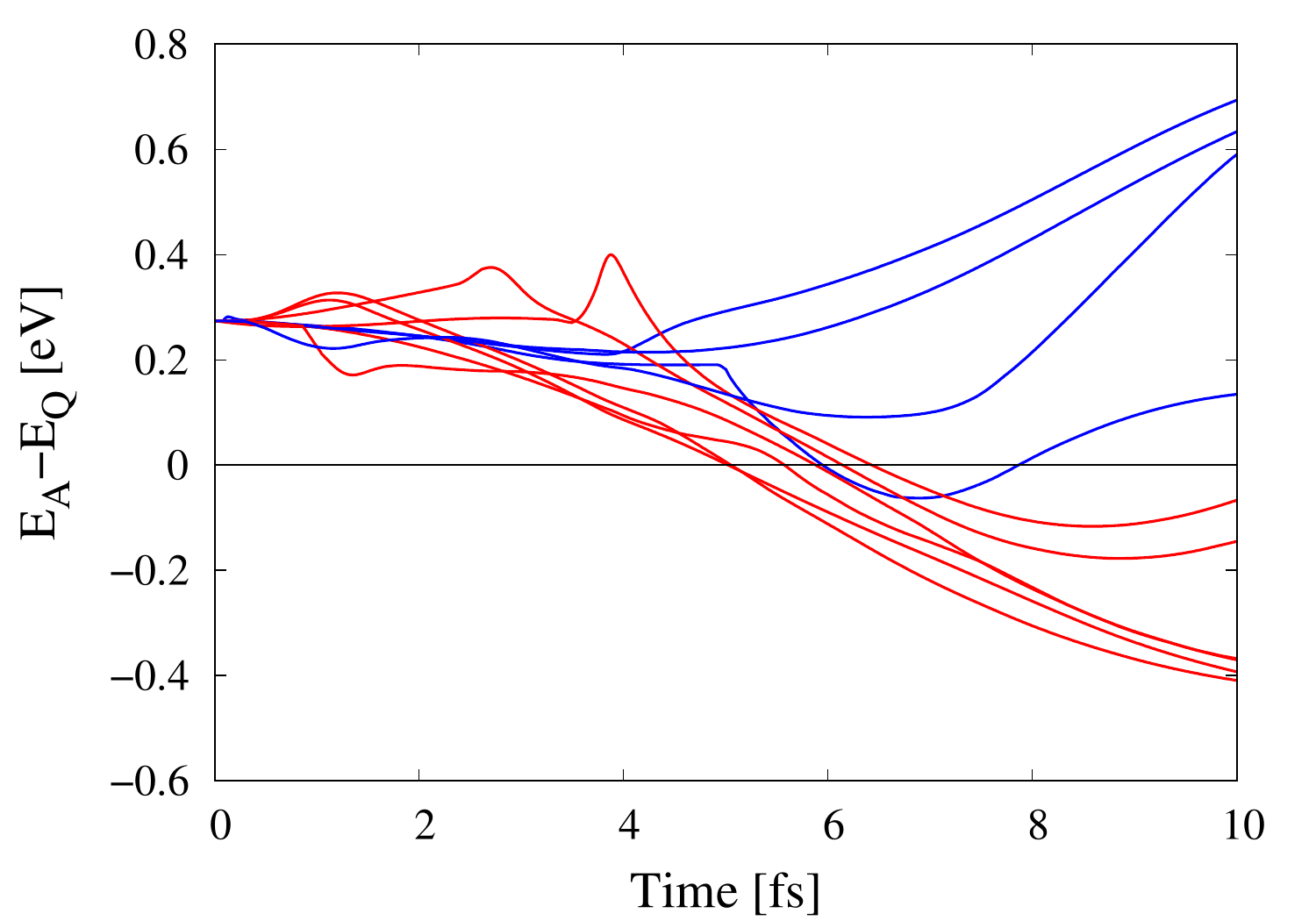}
    \caption{Diabatic energy difference $E_A - E_Q$ at the position of each GBF center during the DD-vMCG simulation of the non-adiabatic dynamics of the FBZ cation upon initial excitation of the pure $\Psi_A$ diabatic state. The color coding is used to highlight similar behavior within the GBF trajectories.\\~}
    \end{minipage}
\end{figure*}

\section*{Modeling of the autocorrelation function for mixed wavepackets with arbitrary weights}

As observed in figure \ref{interference_auto_SI} (also fig.\,5 of the main manuscript), the diabatic components of the full autocorrelation function upon initial population of the mixed wavepackets $\frac{1}{\sqrt{2}}(\Psi_Q\pm\Psi_A)$ correspond almost exactly to the (rescaled) full autocorrelation functions observed upon initial population of the corresponding pure diabatic state. On the other hand figure \ref{init_phase_SI} showed the independence of the autocorrelation function with respect to the initial phase between the two diabatic states. 

On the basis of these observations, we propose to model the time evolution of the full autocorrelation function of any mixed wavepacket using the following expression:
\beq
\mathcal{O}^\mathcal{P}(t) = \mathcal{O}_\tsc{q}^\mathcal{P}(t) + \mathcal{O}_\tsc{a}^\mathcal{P}(t) \approx \mathcal{P} \times \mathcal{O}^\tsc{q}(t) + ( 1-\mathcal{P} ) \times \mathcal{O}^\tsc{a}(t),
\eeq
where $\mathcal{P} \in [0;1]$ is the mixing parameter associated with a generic initial wavepacket (here assumed real valued) of the following form,
\beq
\Psi^\mathcal{P}(0) = \sqrt{\mathcal{P}} \Psi_Q \pm \sqrt{(1-\mathcal{P})} \Psi_A.
\eeq
From this it comes that
\beq
| \mathcal{O}^\mathcal{P}(t) |^2 \approx \Big( \mathcal{P} \, |\mathcal{O}^\tsc{q}(t) |\Big)^2 + \Big( (1-\mathcal{P}) \, |\mathcal{O}^\tsc{a}(t) | \Big)^2 + \mathcal{P} \times (1-\mathcal{P}) \times | \mathcal{O}^\tsc{q}(t) | \times | \mathcal{O}^\tsc{a}(t) | \times 2  \cos\big( \Delta \bar{\varphi} (t)\big),
\label{O_model}
\eeq
where 
\beq
\Delta \bar{\varphi} (t) = \mathrm{arg}\big(\mathcal{O}^\tsc{q}(t)\big)-\mathrm{arg}\big(\mathcal{O}^\tsc{a}(t)\big).
\eeq

\begin{figure}[h]
\includegraphics[scale=0.28, angle=-90]{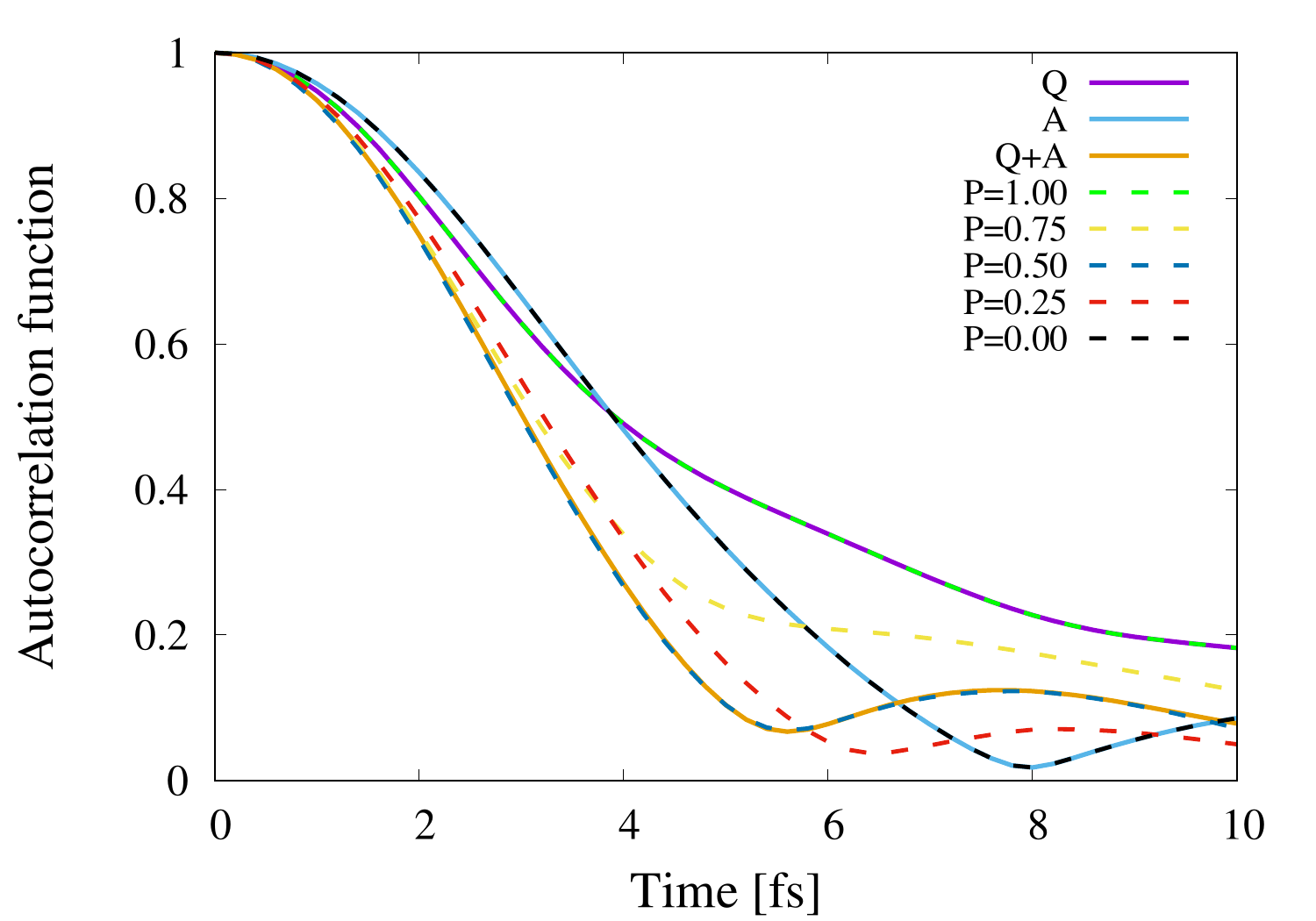}
\includegraphics[scale=0.33, angle=-90, trim= 1.5cm 0 0cm 0, clip]{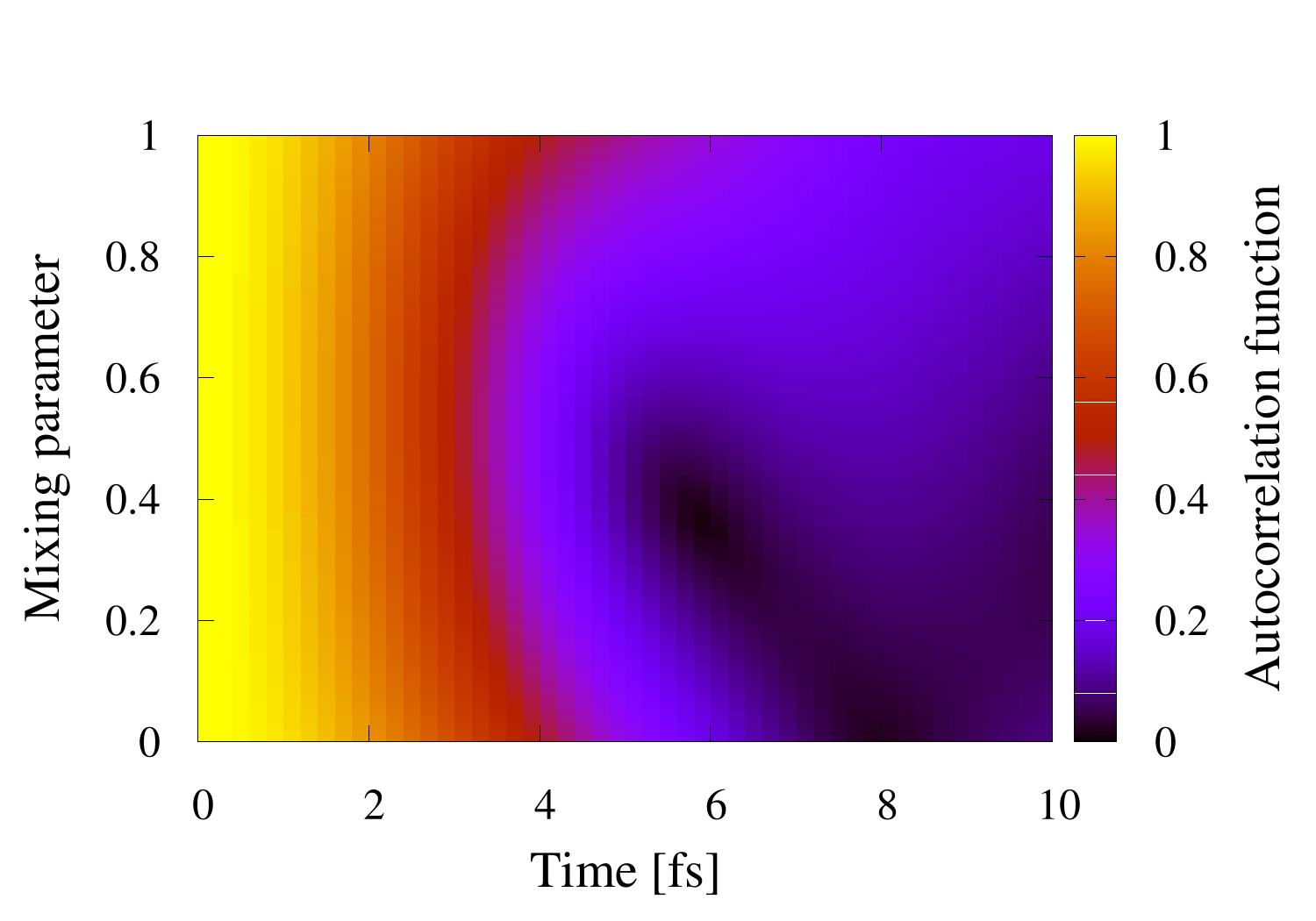}
\caption{\textbf{Left --} Modulus of the autocorrelation functions obtained with DD-vMCG for the initial wavepackets $\frac{1}{\sqrt{2}}(\Psi_Q + \Psi_A)$ (plain orange), $\Psi_Q$ (plain purple) and $\Psi_A$ (plain light blue) and using the extrapolation procedure for selected $\mathcal{P}$ values (dashed curves). \textbf{Right --} Modulus of the extrapolated autocorrelation functions over the full range of $\mathcal{P}$ values.}
\label{map_extrapolated}
\end{figure}

In figure \ref{map_extrapolated} (left) we compare the modulus of the autocorrelation functions obtained using the extrapolation model of equation \eqref{O_model} with the modulus of the autocorrelation functions obtained from our DD-vMCG simulations with the $\frac{1}{\sqrt{2}}(\Psi_Q + \Psi_A)$ initial wavepacket (equivalent to $\mathcal{P}=0.5$) and with the pure initial wavepacket $\Psi_Q$ $(\mathcal{P}=1)$ and $\Psi_A$ $(\mathcal{P}=0)$. By construction the extrapolation method proposed here is equivalent to the exact DD-vMCG result for $(\mathcal{P}=1)$ and $(\mathcal{P}=0)$. As for intermediate value, one can clearly observe that the correspondence is excellent between the extrapolated $| \mathcal{O}^{0.5}(t) |$ and the $|\mathcal{O}^\tsc{q+a}(t)|$ obtained via DD-vMCG. In figure \ref{map_extrapolated} (right) we also show the evolution of the extrapolated $| \mathcal{O}^{\mathcal{P}}(t) |$ as a function over the full range of mixing parameter.

\begin{figure}[h]
\includegraphics[scale=0.45, angle=-90]{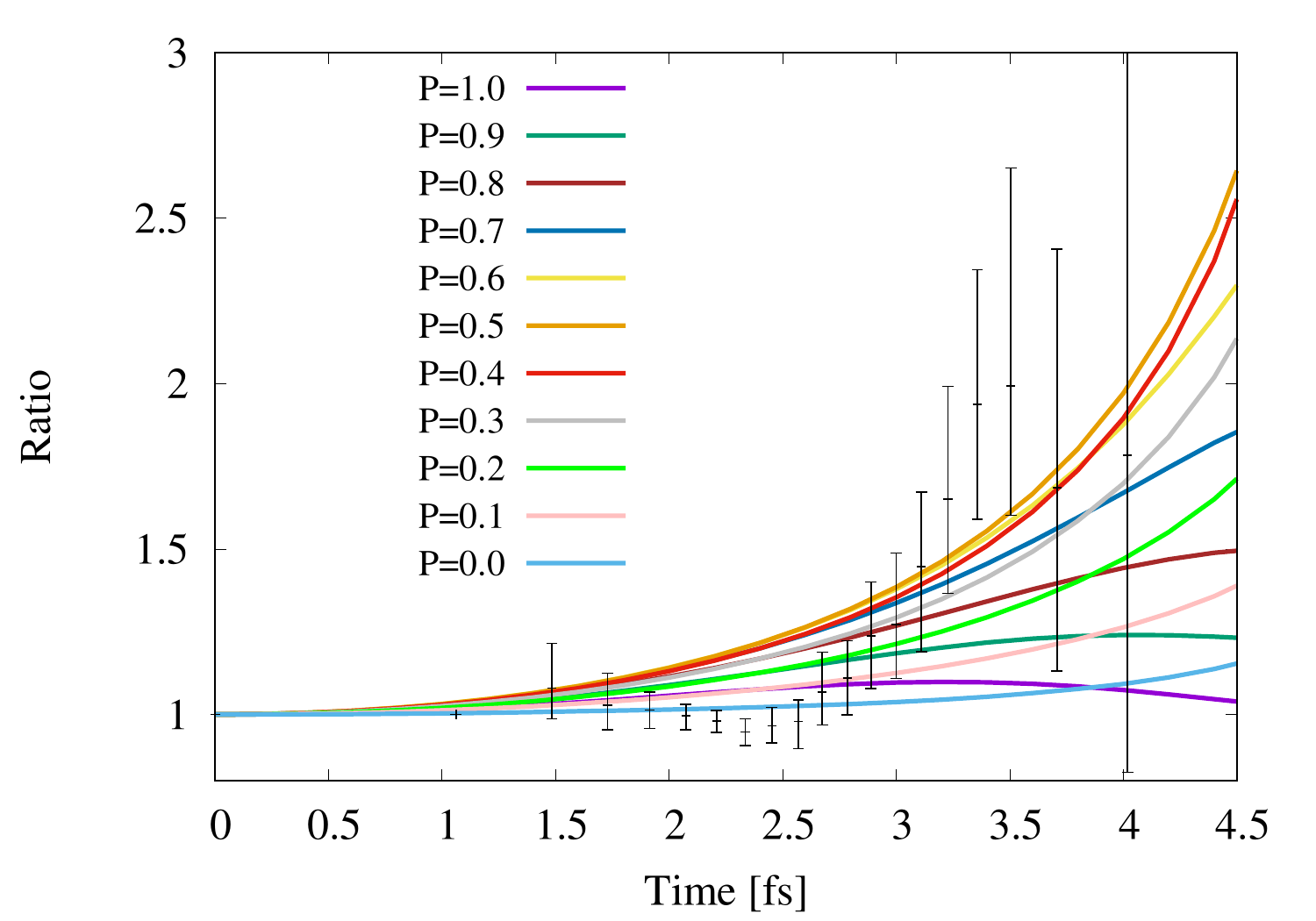}
\caption{Ratio (BZ/FBZ) of experimental $C$ factors (black points) and of the modulus  of the extrapolated autocorrelation functions for various mixing $\mathcal{P}$ values (solid colored lines).}
\label{fig_ratio_extrapolated}
\end{figure}

We now use this model to simulate the ratio of autocorrelation functions between BZ and FBZ for different values of $\mathcal{P}$. It is to be noted that, as the autocorrelation function of BZ is independent of the initial composition of the wavepacket (see figure \ref{DBZ_SI}), no extrapolation is needed in this case. In figure \ref{fig_ratio_extrapolated} this ratio of autocorrelation function modulus for a series of $\mathcal{P}$ value is compared to the ratio of the experimental $C$ factor (see Fig 4 of the main text for the comparison with the DD-vMCG results). The experimental error bars are determined by bootstrapping the 3-component fit procedure over multiple repeated datasets, and represent the 10-90\% percentile intervals. The errors bars on the ratio of the $C$ factors are then obtained using standard error propagation.
As clearly observed here, the fast increase the $C$ factor ratio corresponds relatively well to the behaviour of the autocorrelation function ratio for $\mathcal{P}$ values close to $0.5$. Such observation is coherent with the experimental procedure employed in which no specific orientation of the molecular system was enforced. Considering that at the ground state geometry both cationic states are very close in energy, it is therefore not surprising that both states were populated with almost equivalent weight. However, while this figure clearly illustrates the expected sensitivity of this observable with respect to the relative population of the two diabatic states, it also illustrates that the current experimental signal to noise ratio does not allow us to draw any decisive conclusions as to the composition of the initial wavepacket or the ensemble of initial wavepackets. Indeed, the width of the experimental error bars could be due to limited experimental signal to noise ratio and/or an ensemble of molecules excited to different electronic wavepackets, thus leading to autocorrelation functions decaying at different rates. Performing an experiment with aligned molecules may help disentangling these two factors.

\clearpage